\documentclass[twocolumn,showpacs,preprintnumbers,amsmath,amsfonts,amssymb,aps,prb]{revtex4}

\usepackage{graphicx}
\usepackage{dcolumn}
\usepackage{bm}

\newcommand{\mibs}[1]{\mbox{\scriptsize{${\bf #1}$}}}
\newcommand{\citePRSA}[3]{Proc.\ Roy.\ Soc.\ A {\bf #1}, #3 (#2)}
\newcommand{\citePR}[3]{Phys.\ Rev. {\bf #1}, #3 (#2)}
\newcommand{\citePRB}[3]{Phys.\ Rev.\ B {\bf #1}, #3 (#2)}
\newcommand{\citePRL}[3]{Phys.\ Rev.\ Lett. {\bf #1}, #3 (#2)}
\newcommand{\citeRMP}[3]{Rev.\ Mod.\ Phys. {\bf #1}, #3 (#2)}
\newcommand{\citeJPSJ}[3]{J.\ Phys.\ Soc.\ Jpn. {\bf #1}, #3 (#2)}
\newcommand{\citePTP}[3]{Prog.\ Theor.\ Phys. {\bf #1}, #3 (#2)}
\newcommand{\citeJPCS}[3]{J.\ Phys.\ Chem.\ Sol. {\bf #1}, #3 (#2)}
\newcommand{\citeEPJB}[3]{Eur.\ Phys.\ J. B {\bf #1}, #3 (#2)}
\newcommand{\citeJPF}[4]{J.\ Phys.\ (France) #1 {\bf #2}, #4 (#3)}
\newcommand{\citeIBID}[3]{ibid. {\bf #1}, #3 (#2)}

\begin{document}

\title{Ginzburg-Landau Theory and Classical Critical Phenomena of Mott Transition}

\author{Shigeki Onoda}

\affiliation{
Tokura Spin Superstructure Project, ERATO, Japan Science and Technology Agency, \\
c/o Department of Applied Physics, University of Tokyo, Hongo 7-3-1, Tokyo 113-8656, Japan}

\date{Received}

\begin{abstract}
Theory of classical critical phenomena of Mott transition is developed for the dimensionality $d \le \infty$. Reconsidering a cluster dynamical mean-field theory (DMFT), Ginzburg-Landau free energy is derived in terms of hybridization function for a cluster-impurity model. Its expansion around a cluster DMFT solution reduces to a $\phi^4$ model. Inherent thermal Mott transition without spontaneous symmetry breaking is described by a scalar field reflecting the charge degrees of freedom. In the space of local Coulomb repulsion, chemical potential and temperature, a first-order transition surface terminates at a critical end curve. The criticality belongs to the Ising universality as a liquid-gas transition. Various quantities including double occupancy, electron filling and entropy show diverging responses at the criticality and discontinuities at the first-order transition. Particularly, electron effective mass shows a critical divergence in $2\le d\le 4$. Only at a certain curve on the surface, a filling-control transition and its related singularities disappear. We discuss detailed critical behaviors, effects of interplay with other critical fluctuations, and relevant experimental results.
\end{abstract}


\maketitle


\section{Introduction}
\label{sec:Introduction}

Spontaneous symmetry breaking occurs in a variety of phase transitions in correlated electron systems. In particular, at partial but commensurate electron fillings $n$, some of them appear as a metal-insulator transition (MIT). For an integer $n$, the effect is most pronounced, because the MIT in this case is triggered by severe competition between kinetic energy and {\it local} Coulomb repulsion $U$ for localized orbitals. This MIT is classified into a prototype of Mott transition~\cite{Mott}. Here, translational invariance is partly broken and a unit cell is enlarged. This results in a reconstruction of the bands which are either fully occupied or empty in the reduced first Brillouin zone, and hence an insulator explained by Slater~\cite{Slater}. 

\begin{figure}[htb]
\begin{center}
\includegraphics[width=7.5cm]{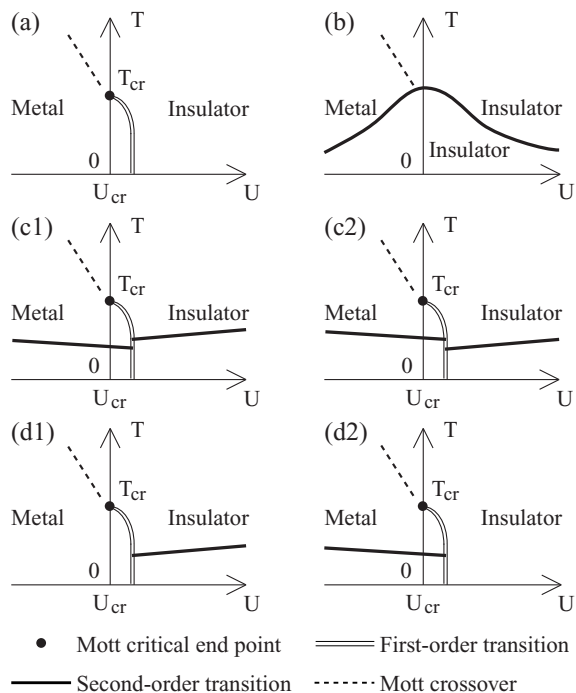}
\caption{Schematic phase diagrams for (a) inherent (liquid-gas) thermal Mott transition and a related crossover, and (b-d) the interplay with other symmetry-breaking phase transitions. (b) The inherent Mott transition is masked by the Slater-type one. (c1/c2) There occur two additional phases separated by the first-order transition from each other, which extends over as an inherent Mott transition at high temperatures. (d1/d2) In the metallic/insulating side, there occurs a symmetry-breaking phase transition. Fluctuation-induced first-order transitions possibly appear in the vicinity of meeting points of these curves~\cite{OnodaNagaosa03}.} 
\label{fig:phasediagram}
\end{center}
\end{figure}

Actually, Mott transition takes place with or without symmetry breaking. It was postulated as a phase transition that occurs at partial but integer $n$ from metal to (magnetically) long-range ordered or disordered insulator with local spin moments \cite{Mott}. A central issue in this paper is the latter prototype of Mott transition occurring without any spontaneous symmetry breaking, which is henceforth referred to as {\it inherent Mott transition}; a first-order Mott transition curve separating the metallic and the insulating states terminates at a critical end point, leaving a crossover at higher temperatures, as shown in Fig.~\ref{fig:phasediagram}~(a). This insulating state is realized by a Hubbard-band splitting due to a large value of $U$ compared with a bandwidth $W$~\cite{Hubbard}. The first-order nature may be assigned to electron correlations even without lattice degrees of freedom. An analogy of the inherent Mott transition to a liquid-gas phase transition~\cite{RMP_Kadanoff} was first discussed by Cyrot~\cite{Cyrot72} and Castellani {\it el al.}~\cite{Castellani79} and demonstrated by a number of experiments on $d$-electron systems such as V$_2$O$_3$ and NiS$_2$~\cite{Mott,RMP_ImadaFujimoriTokura} as well as molecular materials $\kappa$-(BEDT-TTF)$_2$X~\cite{OnodaNagaosa03}.

In the proximity of Mott transition, there appear important phenomena like high-$T_{\rm c}$ cuprate \cite{BednortzMuller86} or organic superconductivity (SC) and possibly spin liquid~\cite{Anderson73} in the resonating valence bond state \cite{RVB}, in addition to antiferromagnetism (AF) due to the superexchange mechanism within a Mott insulating state. Figures~\ref{fig:phasediagram}~(b-d) show schematic phase diagrams for interplay of the inherent Mott transition with other phase transitions, which will also be analyzed in this paper. In (b), the inherent Mott transition is replaced by the Slater's mechanism with a symmetry-breaking Mott transition which has a larger scale of temperature $T$ than the Mott critical temperature $T_{\rm cr}$. In (c) and (d), below $T_{\rm cr}$, there occurs interplay of the inherent Mott transition with other symmetry-breaking transitions that do not appear as an MIT. These phase diagrams are relevant to (c) NiS$_{2-x}$Se$_x$ and $R$NiO$_3$ ($R$: La to Lu)~\cite{Mott,RMP_ImadaFujimoriTokura} for interplay with AF and $\kappa$-(BEDT-TTF)$_2$X for interplay with SC and AF~\cite{OnodaNagaosa03}, (d1) V$_{2-x}$Cr$_x$O$_3$~\cite{Mott,RMP_ImadaFujimoriTokura} and $\kappa$-(BEDT-TTF)$_2$X (under a magnetic field)~\cite{Kagawa_unpublished} for interplay with AF, and (d2) $\kappa$-(BEDT-TTF)$_2$Cu$_2$(CN)$_3$~\cite{ShimizuMiyagawaKanodaMaesatoSaito03} or $R_2$Mo$_2$O$_7$ ($R$: Nd to Y)~\cite{Tokura_MIT_pyrochlore} for interplay with SC or ferromagnetism.

On the paramagnetic metallic side of the Mott transition, quasiparticle mass and hence a specific heat coefficient $C/T$ are enhanced and even diverge at quantum Mott criticality~\cite{BrinkmanRice70}, as obtained in infinite dimensions $d=\infty$~\cite{RMP_DMFT}. The quantum Mott transition for the single-band Hubbard model was described as a holon-doublon binding-unbinding transition~\cite{KotliarRuckenstein86}. Holons and doublons are bound by a large value of $U/W$ at half filling, and then no longer propagate freely, yielding a Mott insulating state. The binding energy gives a Mott gap and thus the inverse of the localization length $\xi_{\rm loc}$. This $\xi_{\rm loc}$ corresponds to the holon-holon (or doublon-doublon) and the holon-doublon coherence lengths. However, understanding critical properties of Mott transitions remains open in realistic spatial dimensions, whether it is quantum or classical. Now, it is important to clarify their universal properties.

Second-order and some first-order classical phase transitions can be understood by analyzing Ginzburg-Landau (GL) free energy. We can even describe critical phenomena of conventional second-order ones in terms of universality class uniquely specified by the number of components of relevant order parameters in each system dimension, as far as the interaction is short-range~\cite{Stanley87}. Then, it is fundamental to find a minimal complete set of relevant order parameters. For symmetry-breaking phase transitions, one may be ready to select the order parameters conjugate to the symmetry-breaking fields and hence to construct the GL free energy. In contrast, for liquid-gas phase transitions occurring without any symmetry breaking, the two states on the both sides of the first-order phase transition are adiabatically connected. The relevant order parameter may be hidden, and hence it is nontrivial to derive the GL theory. 

Only in $d=\infty$, GL theory of Mott transition has been developed by Kotliar {\it el al.}~\cite{GL_DMFT} along the formalism of the dynamical mean-field theory (DMFT)~\cite{RMP_DMFT,DMFA}. ``Order parameter'' is derived from hybridization function in the single-impurity Anderson model as a bosonic scalar field. Its finite-temperature criticality belongs to the mean-field (MF) universality. This GL theory together with DMFT solutions of the Hubbard model implies that the doublon and the charge susceptibilities
\begin{eqnarray}
  \chi_{DD}&=&-\frac{1}{N}\frac{\partial^2 F}{\partial U^2}
  \label{eq:chi_DD_def}\\
  \chi_{nn}&=&-\frac{1}{N}\frac{\partial^2 F}{\partial \mu^2}
  \label{eq:chi_nn_def}
\end{eqnarray}
simultaneously diverge at Mott critical end curve in the space of $(U,\mu,T)$~\cite{KotliarMurthyRozenberg02} where $F$, $\mu$ and $N$ are a free energy for the grand canonical ensemble, a chemical potential and the number of atomic sites, respectively. However, detailed analyses in the whole $(U,\mu,T)$ space remain open. Besides, in finite dimensions $d<\infty$, even derivation of GL free energy is missing.

In this paper, we derive GL free energy of Mott transition for $d \le \infty$. GL free energy describing the inherent thermal Mott transition turns out to be similar to that of the Ising model under a magnetic field. Then, we confirm that the analogy to the liquid-gas phase transition~\cite{RMP_Kadanoff} holds, but with {\it crucial modifications of the DMFT results by spatial correlations}~\cite{OnodaNagaosa03} in $2\le d \le d_c$ where $d_c=4$ is the critical dimension~\cite{Goldenfeld}. This is consistent with the absence of thermal Mott transition for $d=1$, a phase diagram proposed for the Hubbard model in $d=2$~\cite{CPM}, and a recent experimental scaling analysis of the resistivity in V$_{2-x}$Cr$_x$O$_3$~\cite{Science_Limelette03}.

\begin{figure}
\begin{center}
\includegraphics[width=7.5cm]{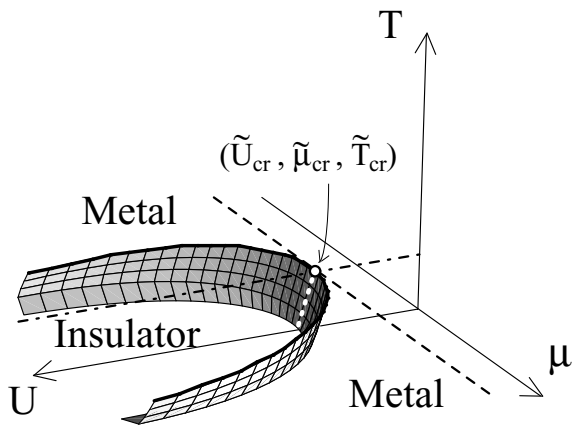}
\end{center}
\caption{Schematic phase diagram of the inherent Mott transition in the ($U$,$\mu$,$T$) space. A first-order transition surface (shaded surface) terminates at a critical end curve (thick solid curve). Within the surface, there exists a curve (open dotted curve) where a first-order bandwidth-control transition with a discontinuity of $D$ occurs but not a filling-control. This special curve terminates at a critical end point $(\tilde{U}_{\rm cr},\tilde{\mu}_{\rm cr},\tilde{T}_{\rm cr})$ (open circle) on the critical end curve. At this point, a second-order bandwidth-control transition with $\chi_{DD}\to\infty$ occurs. Except at the special curve and point, both $D$ and $n$ show discontinuities at the first-order transition surface and their $U$ and $\mu$ derivatives diverge at the critical end curve. The dashed line represents $U=\tilde{U}_{\rm cr}$ and $T=\tilde{T}_{\rm cr}$, while the dash-dotted line  $\mu=\tilde{\mu}_{\rm cr}$ and $T=\tilde{T}_{\rm cr}$.}
\label{fig:phasediagram2}
\end{figure}

We propose a $(U,\mu,T)$ phase diagram for the inherent Mott transition (Fig.~\ref{fig:phasediagram2}). A first-order Mott transition surface (shaded area) terminates at a critical end curve (thick solid curve). Both bandwidth-control and filling-control Mott transitions occur there, except at a special curve (thick dotted curve) and its end point $(\tilde{U}_{\rm cr},\tilde{\mu}_{\rm cr},\tilde{T}_{\rm cr})$ (open circle) where a bandwidth-control transition takes place but not a filling-control with $U$ being fixed. The double occupancy $D$ is discontinuous at the first-order transition surface and $\chi_{DD}$ diverges at the critical end curve. The electron filling $n$ also shows a discontinuity at the surface but at the special bandwidth-control curve where its singular component vanishes. The same critical divergence as $\chi_{DD}$ emerges for $\chi_{nn}$ and 
\begin{equation}
  \chi_{Dn}=-\frac{1}{N}\frac{\partial^2 F}{\partial \mu \partial U}
  \label{eq:chi_Dn_def}
\end{equation}
at the critical end curve but at $(\tilde{U}_{\rm cr},\tilde{\mu}_{\rm cr},\tilde{T}_{\rm cr})$ where diverging components of $\chi_{nn}$ and $\chi_{Dn}$ vanish. $\chi_{Dn} \to \infty$ at criticality is a quite new feature. These reflect that each of $D$ and $n$ contains a linear term in the relevant scalar field. Variations of $U$, $\mu$ or $T$ merely change the expansion coefficients of the GL free energy. Taking into account this aspect, we have succeeded in obtaining a unified picture of the Mott transitions with respect to different control parameters. 

In Sec.~\ref{sec:CDMFT}, to establish a basis of GL arguments for Mott transition, we reconsider a general theoretical framework of a cluster extension of the DMFT.

In Sec.~\ref{sec:GL}, GL free energy functional for this cluster DMFT is constructed in terms of hybridization function of a {\it cluster-impurity} model. In the limit of an infinite cluster size, expanding the free energy around a critical point and transforming fermionic field to bosonic, we obtain a $\phi^4$ model for the GL free energy.

In Sec.~\ref{sec:Mott_liquid-gas}, we consider the inherent Mott transition with finite-temperature criticality. We reveal universal properties of various physical quantities, the connection among Mott transitions driven by different control parameters, and the phase diagram shown in Fig.~\ref{fig:phasediagram2}.

Interplay of the inherent Mott transition with other symmetry-breaking transitions is a topic of Sec.~\ref{sec:interplay}. From GL arguments, as far as critical regions of the other transitions do not overlap, the Mott criticality appears at an end point of the liquid-gas phase transition or is masked by the other phase transition, as shown in Fig.~\ref{fig:phasediagram}. On the other hand, one-loop renormalization group analyses show that in the presence of the appreciable overlap, it is replaced with a fluctuation-induced first-order transition, as in a previous discussion on (c) for interplay with SC and AF~\cite{OnodaNagaosa03}.

Lastly, we summarize the present results and comment on problems left for future studies in Sec.~\ref{sec:Summary}.


\section{Cluster Dynamical Mean-Field Theory}
\label{sec:CDMFT}

There have been attempts to restore spatial correlations ignored in the DMFT~\cite{CPM,DCA,CDMFT}. On one hand, correlator projection method~\cite{CPM} combines the DMFT with projection methods~\cite{OPM,MatsumotoMancini97,AvellaManciniMunzner01} of electronic operators leading to continued-fraction expansion of Green's functions in energy $\omega$~\cite{OP}. On the other hand, cluster approaches~\cite{DCA,CDMFT} solve a cluster problem containing multiple impurities instead of a single impurity in the DMFT. They should become exact if the cluster size is taken infinite. However, depending on the formalisms, Green's functions and self-energy parts have different properties. In this section, we reexamine a general framework of the cluster DMFT, which gives a basis for constructing the GL free energy.

\subsection{General requirements}
\label{subsec:CDMFT:requirements}

We impose the following three requirements, whichever the cluster is introduced in a real or a momentum space.

(i) Dynamical quantities such as Green's function and self-energy parts are manifestly causal. Difficulty in satisfying this has already been overcome in the literature~\cite{DCA,CDMFT}.

(ii) Each symmetry of the Hamiltonian is spontaneously broken in a solution if and only if order parameter conjugate to the symmetry breaking field remains finite, except the translational invariance which is not necessarily guaranteed because each cluster is treated separately from the others. This requirement is crucial for classifying phase transitions.

(iii) Self-energy parts including the secular MF (Hartree-Fock and/or BCS) terms are only taken into account within the cluster. This facilitates arguments to derive a cluster DMFT equation. Although this requirement introduces an unphysical barrier between the cluster and the exterior medium, this boundary effect is $O(L_c^{d-1})$ for the cluster size $N_c=L_c^d$ and hence negligible compared to $L_c^{d}$ as $L_c\to\infty$ at least for $d\ge2$.

\subsection{Lattice model}
\label{subsec:CMDFT:lattice}

To be explicit, we consider an $M$-component electron system described by the following Hamiltonian;
\begin{eqnarray}
  {\cal H} &=& \sum_{\mibs{k},\sigma_1,\sigma_2}
  \varepsilon_{\sigma_1,\sigma_2}({\bf k})
  c^\dagger_{\mibs{k},\sigma_1}
  c_{\mibs{k},\sigma_2}
  \nonumber\\
  &&+\sum_{\mibs{k}_1,\cdots,\mibs{k}_4}
  \sum_{\sigma_1,\cdots,\sigma_4}
  V_{\sigma_1,\sigma_2;\sigma_3,\sigma_4}
  ({\bf k}_1,{\bf k}_2;{\bf k}_3,{\bf k}_4)
  \nonumber\\
  &&\hspace{25mm}\times
  c^\dagger_{\mibs{k}_1,\sigma_1}
  c^\dagger_{\mibs{k}_2,\sigma_2}
  c_{\mibs{k}_3,\sigma_3}
  c_{\mibs{k}_4,\sigma_4}.
\label{eq:H}
\end{eqnarray}
Here, $c_{\mibs{k},\sigma}$ and $c_{\mibs{k},\sigma}^\dagger$ are annihilation and creation operator of electron with the momentum ${\bf k}$ and the internal component degrees of freedom $\sigma$. For instance, $\sigma$ represents spins, orbitals and sublattices. $V$ represents a short-range Coulomb interaction for the electrons and is finite only when ${\bf k}_1+{\bf k}_2={\bf k}_3+{\bf k}_4$. $\varepsilon_{\sigma_1,\sigma_2}({\bf k})$ is given by
\begin{equation}
  \varepsilon_{\sigma_1,\sigma_2}({\bf k}) 
  = -t_{\sigma_1,\sigma_2}({\bf k})-\mu\delta_{\sigma_1,\sigma_2}
  -h_{\sigma_1,\sigma_2}.
  \label{eq:eps}
\end{equation}
$t_{\sigma_1,\sigma_2}({\bf k})$ is an electron transfer with the momentum ${\bf k}$ from $\sigma_2$ to $\sigma_1$ and satisfies $\sum_{\mibs{k}}t_{\sigma_1,\sigma_2}({\bf k})=0$. $\mu$ denotes a chemical potential, and $h_{\sigma_1,\sigma_2}$ represents a traceless uniform field which is locally coupled to electrons and breaks a symmetry in the component degrees of freedom, e.g., magnetic field, energy-level splitting among different orbitals, lattice distortion in the Jahn-Teller term, etc. 

We define electron thermal Green's function at inverse temperature $\beta=1/T$ as
\begin{equation}
G_{\sigma,\sigma'}(i\omega_n;{\bf k},{\bf k}') \equiv -\int_0^\beta\!d\tau\ 
\langle [ c_{\mibs{k},\sigma}(\tau), c^\dagger_{\mibs{k}',\sigma'}]_+ \rangle e^{i\omega_n\tau}
\label{eq:G}
\end{equation}
with $[A,B]_+\equiv AB+BA$ and $c_{\mibs{k},\sigma}(\tau)\equiv e^{{\cal H}\tau}c_{\mibs{k},\sigma}e^{-{\cal H}\tau}$. Henceforth, $\omega_n=(2n+1)\pi T$ and $\Omega_m=2m\pi T$ denote fermionic and bosonic Matsubara frequencies, respectively. The Dyson equation takes a matrix form;
\begin{equation}
  \hat{G}^{-1}(i\omega_n;{\bf k},{\bf k}')
  =\left[i\omega_n\hat{I}-\hat{\varepsilon}({\bf k})\right]\delta_{{\bf k},{\bf k}'}-\hat{\Sigma}(i\omega_n;{\bf k},{\bf k}').
  \label{eq:Dyson}
\end{equation}
We have introduced notations $\hat{G}$, $\hat{\Sigma}$, $\hat{I}$ and $\hat{\varepsilon}$ for $M \times M$ matrices $G_{\sigma,\sigma'}$, $\Sigma_{\sigma,\sigma'}$, $\delta_{\sigma,\sigma'}$ and $\varepsilon_{\sigma,\sigma'}$~\cite{note_superconductivity}, respectively, where $\Sigma_{\sigma,\sigma'}(i\omega_n;{\bf k},{\bf k}')$ is the self-energy part.

\subsection{Cluster model}
\label{subsec:CDMFT:cluster}

To calculate $\hat{\Sigma}(i\omega_n;{\bf k},{\bf k}')$, we introduce as a representative of the original problem a cluster model consisting of $N_c$ sites in the real space as in the cellular DMFT \cite{CDMFT} or $N_c$ momentum cells as in the dynamical cluster approximation (DCA) \cite{DCA}. We define the Green's function $\hat{G}_c(i\omega_n;i,i')$ and the self-energy part $\hat{\Sigma}_c(i\omega_n;i,i')$ for the cluster model, on which we impose self-consistent conditions between cluster and lattice variables;
\begin{subequations}
\begin{eqnarray}
\hat{\Sigma}(i\omega_n;{\bf k},{\bf k}')&=&\frac{1}{N_c}\sum_{i,i'}R({\bf k},i)\hat{\Sigma}_c(i\omega_n;i,i')R^\dagger(i',{\bf k}')
\nonumber\\
\label{eq:Sigmac2Sigma}\\
\hat{G}_c(i\omega_n;i,i')&=&\frac{1}{N}\sum_{\mibs{k},\mibs{k}'}R^\dagger(i,{\bf k})\hat{G}(i\omega_n;{\bf k},{\bf k}')R({\bf k}',i').
\nonumber\\
\label{eq:G2Gc}
\end{eqnarray}
\label{eq:Sigmac2Sigma,G2Gc}
\end{subequations}
Here, $i$ and $i'$ represent the $N_c$ degrees of freedom of the cluster. $R$ is a transformation matrix from cluster to lattice variables characterizing the way of introducing a cluster. It obeys the orthonormal condition, 
\begin{equation}
  \frac{1}{N} \sum_{\mibs{k}} R^\dagger(i,{\bf k}) R({\bf k},i')
       = \delta_{i,i'}.
       \label{eq:R:orthonormal2}
\end{equation}
If we take $R({\bf k},i)=\sqrt{N_c}$ for ${\bf k}$ belonging to $i$th momentum hypercube centered at ${\bf k}(i)$ and $0$ otherwize, it gives a modification of the DCA~\cite{DCA,noteDCA}. The choice of $R({\bf k},i)=e^{i{\mibs{k}}{\mibs{r}}(i)}$ with ${\bf r}(i)$ being the position of $i$th site within the cluster is taken in the cellular DMFT~\cite{CDMFT}.

The cluster model can be derived by integrating out the remaining degrees of freedom other than the cluster along the concept of the cavity construction~\cite{RMP_DMFT}. However, in finite dimensions $d<\infty$, the exterior degrees of freedom mediate additional one-body and many-body interactions among the cluster degrees of freedom~\cite{DCA,CDMFT}. Neglecting the exterior-mediated many-body interactions, we only include the one-body term arising from the hybridization between the cluster and the exterior as in the literature~\cite{CDMFT,DCA}. 

Thus, we obtain the following path-integral expressions of the partition function and the effective action for the cluster in terms of Grassman field $f$ and $\bar{f}$ representing the electronic degrees of freedom within the cluster,
\begin{subequations}
  \begin{eqnarray}
    Z_c&=&\int\!{\cal D}\bar{f}{\cal D}f\exp\left[-S_c\right]
    \label{eq:Z_c}\\
    S_c&=&-\beta\!\!\!\!\!\sum_{n,\sigma_1,\sigma_2,i_1,i_2}\!\!\!\!\!
    \bar{f}_{i_1,\sigma_1}(i\omega_n)\mathfrak{G}^{-1}_{\sigma_1,\sigma_2}(i\omega_n,i_1,i_2)f_{i_2,\sigma_2}(i\omega_n)
    \nonumber\\
    &&+\int_0^\beta\!d\tau\,V_{\sigma_1,\sigma_2;\sigma_3,\sigma_4}
    (i_1,i_2;i_3,i_4)
    \nonumber\\
    &&\times\bar{f}_{i_1,\sigma_1}(\tau)f_{i_2,\sigma_2}(\tau)
    \bar{f}_{i_3,\sigma_3}(\tau)f_{i_4,\sigma_4}(\tau)
    \label{eq:S_c}
\end{eqnarray}
\label{eq:cluster}
\end{subequations}
with
\begin{eqnarray}
  \lefteqn{V_{\sigma_1,\sigma_2;\sigma_3,\sigma_4}(i_1,i_2;i_3,i_4)}
  \nonumber\\
  &&=\sum_{\mibs{k}_1,\cdots,\mibs{k}_4}
  V_{\sigma_1,\sigma_2;\sigma_3,\sigma_4}
  ({\bf k}_1,{\bf k}_2;{\bf k}_3,{\bf k}_4)
  \nonumber\\
  &&\times R^\dagger({\bf k}_1,i_1)R^\dagger({\bf k}_2,i_2)
  R({\bf k}_3,i_3)R({\bf k}_4,i_4).
  \label{eq:V}
\end{eqnarray}
$\hat{\mathfrak{G}}$ plays a role of the Weiss field acting on the cluster in the original lattice model, and is now nonlocal within the cluster in contrast to the conventional DMFT where it is completely local. $\hat{\mathfrak{G}}$ is obtained from
\begin{equation}
  \hat{\mathfrak{G}}^{-1}(i\omega_n;i,i')
  =i\omega_n\delta_{i,i'}\hat{I}-\hat{\varepsilon}_0(i,i')
  -\hat{\Delta}(i\omega_n;i,i'),
  \label{eq:CI:WeissG_inv}
\end{equation}
where
\begin{equation}
  \hat{\varepsilon}_0(i,i')=\frac{1}{N}\sum_{\mibs{k}}R^\dagger(i,{\bf k})\hat{\varepsilon}({\bf k})R({\bf k},i').
  \label{eq:CI:Eii'}
\end{equation}
is the intra-cluster part of $\hat{\varepsilon}({\bf k})$, and
\begin{equation}
\hat{\Delta}(i\omega_n;i,i')=\frac{1}{N}\sum_{\mibs{k},\mibs{k}'}
  \hat{\varepsilon}^\dagger_1(i,{\bf k})
  \hat{G}_{\rm e}(i\omega_n;{\bf k},{\bf k}')
  \hat{\varepsilon}_1({\bf k}',i')
  \label{eq:CI:Delta}
\end{equation}
is the exterior-mediated dynamical one-body interaction, i.e., the hybridization function in the context of a multi-impurity extension of the Anderson impurity model. Here, we have introduced the Green's function for the original lattice model in the absence of the cluster
\begin{eqnarray}
 \lefteqn{\hat{G}_{\rm e}(i\omega_n;{\bf k},{\bf k}')
    =\hat{G}(i\omega_n;{\bf k},{\bf k}')}
  \nonumber\\
  &&{}-\frac{1}{N^2}\sum_{\mibs{k}_1,\mibs{k}_2}\sum_{\mibs{k}_1',\mibs{k}_2'}\sum_{i,i'}
  \hat{G}(i\omega_n;{\bf k},{\bf k}_1)
  R({\bf k}_1,i)
  R^\dagger(i,{\bf k}_2)
  \nonumber\\
  &&\times 
  \hat{G}^{-1}(i\omega_n;{\bf k}_2,{\bf k}_2')
  R({\bf k}_2',i')
  R^\dagger(i',{\bf k}_1')\hat{G}(i\omega_n;{\bf k}_1',{\bf k}'),
  \nonumber\\
  \label{eq:G_e}
\end{eqnarray}
and hybridizations between the cluster and the exterior
\begin{subequations}
\begin{eqnarray}
  \hat{\varepsilon}_1^\dagger(i,{\bf k})&=&R^\dagger(i,{\bf k})\hat{\varepsilon}({\bf k})-\sum_{i'}\hat{\varepsilon}_0(i,i')R^\dagger(i',{\bf k}),
  \label{eq:CI:Eik}\\
  \hat{\varepsilon}_1({\bf k},i)&=&\hat{\varepsilon}({\bf k})R({\bf k},i)-\sum_{i'}R({\bf k},i')\hat{\varepsilon}_0(i',i).
  \label{eq:CI:Eki}
\end{eqnarray}
\end{subequations}
The second terms in Eqs.~(\ref{eq:CI:Eik}) and (\ref{eq:CI:Eki}) remove the intra-cluster contributions to the hybridizations to avoid a double counting of the cluster degrees of freedom. 

The Green's function $\hat{G}_c$ and the self-energy part $\hat{\Sigma}_c$ for the cluster are obtained by solving the cluster model Eq.~(\ref{eq:cluster}) and are related through the Dyson equation
\begin{equation}
\hat{G}_c^{-1}(i\omega_n;i,i')=\hat{\mathfrak{G}}^{-1}(i\omega_n;i,i')-\hat{\Sigma}_c(i\omega_n;i,i').
\label{eq:Dyson_c}
\end{equation}

Finally, Eq.~(\ref{eq:G2Gc}) is reproduced by substituting Eq.~(\ref{eq:CI:Delta}) together with Eqs.~(\ref{eq:G_e}) and (\ref{eq:Sigmac2Sigma}) into Eq.~(\ref{eq:CI:WeissG_inv}), and then the result into Eq.~(\ref{eq:Dyson_c}). Equation~(\ref{eq:Dyson}) together with Eqs.~(\ref{eq:Sigmac2Sigma,G2Gc}) and (\ref{eq:Dyson_c}) defines a set of cluster DMFT self-consistent equations.

\section{Ginzburg-Landau Free Energy Functional}
\label{sec:GL}

In this section, we construct a GL free energy functional in terms of the hybridization function so that for symmetry-unbroken states on the Bethe lattice in $d=\infty$, it reproduces the Kotliar's result~\cite{GL_DMFT}. In $d=\infty$, the hybridization function was regarded as a metallic order parameter. For $d<\infty$, it is nonlocal within the cluster. Therefore, the {\it cluster-average} low-energy behavior of the trace of the hybridization function plays a role of the metallic order parameter. Then, we define a bosonic field corresponding to the ``order parameter'' and describing the soft mode at criticality, and finally obtain a $\phi^4$ model, which facilitates analyses of the GL free energy as will be done in Secs.~\ref{sec:Mott_liquid-gas} and \ref{sec:interplay}.

\subsection{Construction of free energy functional in terms of hybridization function}
\label{subsec:GL:construction}

As is clear from Eqs.~(\ref{eq:cluster}) and (\ref{eq:CI:WeissG_inv}), the effective action $S_c$ and the partition function $Z_c$ are functionals of $\hat{\Delta}$ in addition to a set of model parameters denoted by $\vec{s}=\{U,t,t',\mu,\hat{h},T,\cdots\}$. Then, the cluster Green's function $\hat{G}_c$ and self-energy part $\hat{\Sigma}_c$ are determined by $\hat{\Delta}$ from
\begin{eqnarray}
  \hat{G}_c[\hat{\Delta},\vec{s}](i\omega_n;i,i') &=& 
  \frac{1}{T}\frac{\delta F_c}{\delta \hat{\Delta}}[\hat{\Delta},\vec{s}](i\omega_n;i,i')
  \label{eq:G_c=dF_c/dDelta}\\
  F_c[\hat{\Delta},\vec{s}] &=& -T\log Z_c[\hat{\Delta},\vec{s}],
  \label{eq:F_c}
\end{eqnarray}
and from Eqs.~(\ref{eq:CI:WeissG_inv}) and (\ref{eq:Dyson_c}), namely, 
\begin{eqnarray}
  \hat{\Sigma}_c[\hat{\Delta},\vec{s}](i\omega_n;i,i')
  &=&i\omega_n\delta_{i,i'}\hat{I}-\hat{\varepsilon}_0(i,i')-\hat{\Delta}(i\omega;i,i')
  \nonumber\\
  &&{}-\hat{G}_c^{-1}[\hat{\Delta},\vec{s}](i\omega_n;i,i'),
  \label{eq:Sigma_c[Delta]}
\end{eqnarray}
respectively. Therefore, the lattice Green's function Eq.~(\ref{eq:Dyson}) with the self-energy part Eq.~(\ref{eq:Sigmac2Sigma}) can be regarded as a functional of $\hat{\Delta}$ as well;
\begin{eqnarray}
  \lefteqn{\hat{G}^{-1}[\hat{\Delta},\vec{s}](i\omega_n;{\bf k},{\bf k}')
  =\left[i\omega_n\hat{I}-\hat{\varepsilon}({\bf k})\right]\delta_{\mibs{k},\mibs{k}'}}
  \nonumber\\
  &&{}-\sum_{i,i'}R({\bf k},i)\hat{\Sigma}_c[\hat{\Delta},\vec{s}](i\omega_n;i,i')R^\dagger(i',{\bf k}').
  \label{eq:Ginv[Delta]}
\end{eqnarray}
Finally, via a Legendre transformation from $\hat{G}_c$ to $\hat{\Delta}$, we can construct GL free energy $F_{\rm GL}$ as a functional of $\hat{\Delta}$;
\begin{widetext}
\begin{equation}
  \delta F_{\rm GL}[\hat{\Delta},\vec{s}]
  = \delta F_c[\hat{\Delta},\vec{s}]
  -\frac{T}{N}\sum_{n,\mibs{k},\mibs{k}',i,i'}{\rm Tr} 
  R^\dagger(i',{\bf k}) 
  \hat{G}[\hat{\Delta},\vec{s}](i\omega_n;{\bf k},{\bf k}')
  R({\bf k}',i)
  \delta\hat{\Delta}(i\omega_n;i,i'),
  \label{eq:deltaF_GL}
\end{equation}
so that the stationary condition directly gives the cluster DMFT equation given by Eq.~(\ref{eq:G2Gc});
\begin{equation}
  \hat{\Gamma}^{(1)}[\hat{\Delta},\vec{s}](i\omega_n;i,i')
  \equiv \frac{1}{T}\frac{\delta F_{\rm GL}}{\delta \hat{\Delta}}[\hat{\Delta},\vec{s}]
  =\hat{G}_c[\hat{\Delta},\vec{s}](i\omega_n;i,i')
  -\frac{1}{N}\sum_{\mibs{k},\mibs{k}'}
  R^\dagger(i,{\bf k})
  \hat{G}[\hat{\Delta},\vec{s}](i\omega_n;{\bf k},{\bf k}')
  R({\bf k}',i')
  =0.
  \label{eq:stationary}
\end{equation}
\end{widetext}
Especially for the Bethe lattice in $d=\infty$, Eq.~(\ref{eq:CI:Delta}) is simplified as $\Delta(i\omega_n)=\tilde{t}^2G_c(i\omega_n)$ with $N_c=1$, $\tilde{t}\equiv\lim_{d\to\infty}t/\sqrt{d}$ and the nearest-neighbor transfer $t$, and hence Eqs.~(\ref{eq:deltaF_GL}) and (\ref{eq:stationary}) reproduce the Kotliar's result~\cite{GL_DMFT};
\begin{equation}
F_{\rm GL}[\Delta,\vec{s}]=F_c[\Delta,\vec{s}]-T\sum_n\Delta(i\omega_n)^2/2\tilde{t}^2.
\label{eq:F_GL_Kotliar}
\end{equation}

It is remarkable that $\hat{\Delta}$ plays a role of an auxiliary field in the present theory. In conventional functional-integral techniques like the Hubbard-Stratonovich (HS) transformation in the Hubbard model, auxiliary fields can be introduced for describing the magnetic order parameter and long-range ordering. There are three important differences between these two representations. 

(i) The saddle-point solution leads to a cluster DMFT in the present case, while it only gives the Slater's MF theory in the conventional functional-integral method. 

(ii) In the present theory, $\hat{\Delta}$ includes the charge degrees of freedom and their instability in addition to the magnetic ones, while random phase approximations (RPA) or $\phi^4$ theory for itinerant magnetism based on a simple HS transformation~\cite{Moriya,Hertz76,Millis93} can not properly describe the charge degrees of freedom when their fluctuations develop near the Mott criticality. 

(iii) In the present theory, the auxiliary field is fermionic and contains momentum-resolved fermionic spectral properties which gives crucial contributions to the momentum dependence of the self-energy correction in low dimensions, while in the above theories for itinerant magnetism, one only deals with an bosonic field which is locally coupled to fermions.

\subsection{Expansion of Ginzburg-Landau free energy functional around a critical point}
\label{subsec:GL:expansion}

In order to obtain GL expansion of the free energy, we assume that in the original model, a criticality appears at a set of model parameters $\vec{s}=\vec{s}_{\rm cr}$ (temperature $T_{\rm cr}$, local Coulomb interaction $U_{\rm cr}$, chemical potential $\mu_{\rm cr}$, etc). We take the limit of an infinite cluster size $N_c\to\infty$ so that $N_c/N$ becomes a finite value less than unity. We further assume a periodic boundary condition. Then, the hybridization function $\hat{\Delta}(i\omega_n,i,i')$ can be expressed in the energy-momentum space $(i\omega_n,{\bf k})$ as $\hat{\Delta}(i\omega_n,{\bf k})$, since due to the translational symmetry, the momentum conservation law as well as the energy conservation holds. (When we consider a translational-symmetry-breaking transition with a unit-cell enlarging, we need to take this enlarged unit cell at the initial stage of Eqs.~(\ref{eq:H}) and (\ref{eq:eps}).) Then, expanding the free energy $F_{\rm GL}[\hat{\Delta},\vec{s}]$ around the cluster DMFT solution $\hat{\Delta}_{\rm cr}$ at the critical point $\vec{s}_{\rm cr}$ in terms of $\delta \hat{\Delta}=\hat{\Delta}-\hat{\Delta}_{\rm cr}$, we obtain
\begin{equation}
  F_{\rm GL} = F_{\rm GL}^{(0)}+F_{\rm GL}^{(1)}+F^{(2)}_{\rm GL}+F^{(3)}_{\rm GL}+F^{(4)}_{\rm GL}+\cdots
  \label{eq:F_GL}
\end{equation}
with a regular part $F^{(0)}_{\rm GL}=F_{\rm GL}[\hat{\Delta}_{\rm cr},\vec{s}]$ and 
\begin{subequations}
\begin{eqnarray}
  F^{(1)}_{\rm GL}
  &=& T\sum_k {\rm Tr}\left(
  \hat{\Gamma}^{(1)}[\hat{\Delta}_{\rm cr},\vec{s}](k)\,\delta\hat{\Delta}(k)\right)
  \label{eq:F_GL:1}
  \\
  F^{(\ell)}_{\rm GL} &=& \frac{T}{\ell}
  \sum_{K_1,\cdots,K_\ell} 
  \Gamma^{(\ell)}_{K_1|\cdots|K_\ell}[\hat{\Delta}_{\rm cr},\vec{s}]
  \prod_{j=1}^\ell \delta \Delta_{K_j}.
  \label{eq:F_GL:ell}
\end{eqnarray}
\end{subequations}
Here $k_j$ and $K_j$ are abbreviations of $(i\omega_{n_j},{\bf k}_j)$ and $(\sigma_j,\sigma_j')\otimes (i\omega_{n_j},{\bf k}_j)$, respectively, and we have introduced
\begin{equation}
  \Gamma^{(\ell)}_{K_1,|\cdots|K_\ell}[\hat{\Delta},\vec{s}]
  \equiv
  \frac{\ell}{(\ell-1)!}\left(\prod_{j=2}^\ell \frac{\delta}{\delta \Delta_{K_j}}\right) \Gamma^{(1)}_{K_1}[\hat{\Delta},\vec{s}]
    \label{eq:F_GL:Gamma}
\end{equation}
for $\ell\ge2$. In principle, this enables us to obtain the GL expansion parameters for explicit models, although, practically, it may not be easy since it requires extrapolations of cluster Green's function, self-energy part and vertex parts into the limit $N_c\to\infty$ with $N_c/N<1$ being fixed.

\subsection{$\phi^4$-model representation}
\label{subsec:GL:phi4}

We label the real eigenvalues of the hermitian matrix $\Gamma^{(2)}_{\sigma_1,\sigma_1'|\sigma_2,\sigma_2'}[\hat{\Delta}_{\rm cr},\vec{s}](k_1|k_2) \equiv \Gamma^{(2)}_{K_1,K_2}[\hat{\Delta}_{\rm cr},\vec{s}]$ as $\{\gamma^{(2)}_\lambda(p)[\vec{s}]\}$ with $\lambda=0,\cdots,M^2-1$ and an abbreviation $p=(i\Omega_m,{\bf p})$, and the corresponding eigenbases as $\{\Psi_{\sigma,\sigma'|\lambda}[\vec{s}](k|p)\}$. Namely,
\begin{eqnarray}
  &&\sum_{\sigma_2,\sigma_2',k_2}
  \Gamma^{(2)}_{\sigma_1,\sigma_1'|\sigma_2,\sigma_2'}[\hat{\Delta}_{\rm cr},\vec{s}](k_1|k_2)
  \Psi_{\sigma_2,\sigma_2'|\lambda}[\vec{s}](k_2|p)
  \nonumber\\
  &&=\Psi_{\sigma_1,\sigma_1'|\lambda}[\vec{s}](k_1|p)\gamma^{(2)}_\lambda[\vec{s}](p).
  \label{eq:Psi-gamma2}
\end{eqnarray}
 We further introduce $\phi_\lambda(p)$ as the component of $\delta \Delta_{\sigma,\sigma'}(k)$ projected onto $\Psi_{\sigma,\sigma'|\lambda}[\vec{s}](k|p)$;
\begin{equation}
  \delta \Delta_{\sigma,\sigma'}(k)
  = \sum_{\lambda,p}
  \Psi_{\sigma,\sigma'|\lambda}(k|p)
  \phi_\lambda(p).
  \label{eq:phi}
\end{equation}
Owing to the energy-momentum conservation, $\Psi_{\sigma,\sigma'|\lambda}[\vec{s}](k|p)$ contains an energy-momentum $k-p=(i\omega_n-i\Omega_m,{\bf k}-{\bf p})$. Namely, the fermionic function $\Delta_{\sigma,\sigma'}$ is expressed as a convolution of the fermionic function $\Psi_{\sigma,\sigma'|\lambda}$ and the bosonic function $\phi_\lambda$.

Substituting Eq.~(\ref{eq:phi}) into Eqs.~(\ref{eq:F_GL:1}) and (\ref{eq:F_GL:ell}), we obtain 
\begin{subequations}
  \begin{eqnarray}
    F^{(1)}_{\rm GL}
    &=& T \gamma^{(1)}_\lambda \phi_\lambda(0)
    \label{eq:F_GL-phi:1}\\
    F^{(2)}_{\rm GL} 
    &=& \frac{T}{2}\sum_{p} \gamma^{(2)}_\lambda(p)
    |\phi_\lambda(p)|^2
    \label{eq:F_GL-phi:2}\\
    F^{(3)}_{\rm GL}
    &=& \frac{T}{3}\sum_{P_1,P_2,P_3}\gamma^{(3)}_{P_1,P_2,P_3}
    \phi_{P_1} \phi_{P_2} \phi_{P_3}
    \label{eq:F_GL-phi:3}\\
    F^{(4)}_{\rm GL} &=& \frac{T}{4}\sum_{P_1,\cdots,P_4}
    \gamma^{(4)}_{P_1,P_2,P_3,P_4}
    \phi_{P_1} \phi_{P_2} \phi_{P_3} \phi_{P_4}
    \label{eq:F_GL-phi:4}
  \end{eqnarray}
  \label{eq:F_GL-phi}
\end{subequations}
where $P_i$ denotes an abbreviation for a set of $\lambda_i$ and $p_i$ and we have also introduced
\begin{eqnarray}
  \gamma^{(1)}_\lambda[\vec{s}] &=& 
  \sum_K\Gamma^{(1)}_K[\hat{\Delta}_{\rm cr},\vec{s}]
  \Psi_{\sigma,\sigma'|\lambda}[\vec{s}](k|0),
  \label{eq:gamma1}\\
  \gamma^{(\ell)}_{P_1,\cdots,P_\ell}[\vec{s}]\!\!
  &=& \!\!\!\!\sum_{K_1,\cdots,K_\ell}\!\!\!\!
  \Gamma^{(\ell)}_{K_1|\cdots|K_\ell}[\hat{\Delta}_{\rm cr},\vec{s}]
  \prod_{j=1}^\ell\Psi_{\sigma_j,\sigma_j'|\lambda_j}[\vec{s}](k_j|p_j),
  \nonumber\\
  \label{eq:gammaell}
\end{eqnarray}
with $\ell=3$ and $4$. The energy-momentum conservation guarantees that $\gamma^{(1)}$ has only a static uniform component and that $\gamma^{(\ell)}_{P_1,\cdots,P_\ell}$ vanishes unless $\sum_{j=1}^\ell p_j=\sum_{j=1}^\ell(i\Omega_{m_j},{\bf p}_j)=0$ for $\ell=3$ and $4$. We note that $\gamma^{(3)}_{P_1,P_2,P_3}$ and $\gamma^{(4)}_{P_1,P_2,P_3,P_4}$ are symmetric under any permutation between $P_j$ and $P_{j'}$, and that $\gamma^{(1)}_{m'}=\gamma^{(3)}_{m'|m''|m''}(p_1,p_2,p_3)=0$ holds in the presence of $O(m)$ symmetry in $\phi_{m'}$ with $m'=m_0,\cdots,m_0+m-1$ and $0\le m''\le M^2-1$. Eqs.~(\ref{eq:F_GL-phi}) together with Eqs.~(\ref{eq:gamma1}) and (\ref{eq:gammaell}) are reduced to a $\phi^4$ model.

Apparently, phase volumes of the components $(\sigma,\sigma')$ and the energy-momentum $(i\omega_n,{\bf k})$ for $\delta\hat{\Delta}(i\omega_n,{\bf k})$ coincide with those of $\lambda$ and $(i\Omega_m,{\bf p})$ for $\phi_\lambda(i\Omega_m,{\bf p})$, respectively, because we have only performed a unitary transformation Eq.(\ref{eq:phi}). Among these degrees of freedom, there exists a unique field $\phi_0(i\Omega_m,{\bf p})$ whose static uniform part need not spontaneously break a symmetry of the Hamiltonian and can be directly related to the total electron filling and the total multiple occupancy. In particular, if the Mott transition occurs without any symmetry breaking in the single-band model, then the component degrees of freedom $(\sigma,\sigma')$ and $\lambda$ disappear in all of Green's functions, hybridization functions and the field $\phi_\lambda(p)$. Therefore, a field intrinsic to Mott transition has only one component, which is referred to as $\varphi(p)\equiv\phi_0(p)$. This result is physically natural since this scalar field represents nothing but the charge degrees of freedom. Other field components describe relative changes for the spin, orbital, intra-sublattice, and/or pairing degrees of freedom.

\section{Classical Critical Phenomena of Inherent Mott Transition}
\label{sec:Mott_liquid-gas}

In this section, we consider the inherent thermal Mott transition in the absence of interplay with other degrees of freedom. The GL free energy in terms of the scalar field $\varphi(p)$ yields the Ising universality for $2 \le d < d_c$ due to critical order parameter fluctuations and the MF universality for $d_c<d$ as in $d=\infty$~\cite{GL_DMFT} where $d_{\rm c}=4$ is the critical dimension. Here, we discuss various universal properties characteristic of the Mott criticality $2\le d\le\infty$. We clarify a $(U,\mu,T)$ phase diagram shown in Fig.~\ref{fig:phasediagram2} and the connection between the bandwidth-control and the filling-control Mott transitions. Experimental tests of observable quantities are also proposed.

\subsection{Ising universality class}
\label{subsec:Mott_liquid-gas:Ising}

It is reasonable to assume that there occurs an inherent thermal Mott transition for $d\ge2$ that does not break any symmetry spontaneously, as in liquid-gas phase transitions. Our classical GL free energy reads 
\begin{equation}
\frac{F_{\rm GL}}{T}\!=\!\!\int\!\!d{\bf x}\big[-h\varphi({\bf x})+\frac{1}{2}(r\varphi({\bf x})^2+({\bf \nabla}\varphi({\bf x}))^2)+\frac{u}{4}\varphi({\bf x})^4\big].
\label{eq:F_GL-eta}
\end{equation}
$h$, $r$ and $u$ represent the first-, the second- and the fourth-order coupling constants, respectively, which are functions of $\vec{s}$. We have omitted the zeroth-order term $F^{(0)}_{\rm GL}$, which is regular and does not alter singular behavior obtained below. The third-order term $F^{(3)}_{\rm GL}$ has been removed by redefining $\varphi(p)$ as
\[
  \phi_0(p) - \sum_{p_1,p_2,p_3}\gamma^{(3)}_{0,0,0}(p_1,p_2,p_3) {\gamma^{(4)}_{0,0,0,0}}^{-1}(p_1,p_2,p_3,p).
\]
Since the dynamics does not alter the universality of thermal transitions and the nonlocality of the quartic coupling $u$ is irrelevant, they have been neglected in Eq.~(\ref{eq:F_GL-eta}).

In this GL free energy, a criticality takes place at $(h,r)=(0,r_{\rm cr}(u))$ where $r_{\rm cr}(u)$ is a counter-term canceling out the renormalization of $u$ to $r$. Then, the criticality belongs to the Ising universality~\cite{OnodaNagaosa03};
\begin{subequations}
  \begin{eqnarray}
    \xi|_{r,h=0} & \propto & |r-r_{\rm cr}(u)|^{-\nu},
    \label{eq:nu}\\
    C_h|_{r,h=0} & \propto & |r-r_{\rm cr}(u)|^{-\alpha}
    \label{eq:alpha}\\
    \langle \varphi \rangle|_{r,h=0} & \propto & (r_{\rm cr}-r)^\beta \quad
    \mbox{ for $r < r_{\rm cr}$},
    \label{eq:beta}\\
    \chi|_{r,h=0}& \propto & |r-r_{\rm cr}(u)|^{-\gamma},
    \label{eq:gamma}\\
    \langle \varphi \rangle|_{r=r_{\rm cr},h} & \propto & {\rm sgn}(h)|h|^{1/\delta},
    \label{eq:delta}\\
    \chi|_{r=r_{\rm cr},h} & \propto & |h|^{-1+1/\delta},
    \label{eq:delta_chi}
  \end{eqnarray}
  \label{eq:scaling}
\end{subequations}
where $\xi$ is the correlation length characteristic of the fluctuations of $\varphi$, $C_h$ is a specific heat at a constant $h$ on the metal-insulator boundary $h=0$, and $\chi\equiv-\frac{1}{N}\partial^2 F/\partial h^2$ is the susceptibility of $\varphi$. Scaling hypothesis gives relations among the exponents; $\alpha=2-d\nu$, $\beta=\nu(d-2+\eta)/2$, $\gamma=\nu(2-\eta)$, and $\delta=(d+2-\eta)/(d-2+\eta)$~\cite{Ma}. In the case of $d=3$, $\nu\sim0.64$, $\alpha\sim0.09$, $\beta\sim0.33$, $\gamma\sim1.25$, and $\delta\sim4.8$. In the case of $d=2$, $\nu=1$, $\alpha=0$, $\beta=1/8$, $\gamma=7/4$, and $\delta\sim15$. Furthermore, varying $h$ across the Mott metal-insulator boundary $h=0$ gives a first-order transition for $r<r_{\rm cr}(u)$ and a crossover for $r>r_{\rm cr}(u)$. However, when we consider the scaling properties in the $\vec{s}$ space, the absence of the inversion symmetry $\varphi\to-\varphi$ bears complicated properties as in liquid-gas phase transitions~\cite{RMP_Kadanoff,Ma}, which will be discussed later.

\subsection{Phase diagram}
\label{subsec:Mott_liquid-gas:phase-diagram2}

Next, we discuss universal behaviors in terms of the original model parameters $\vec{s}$. Mott metal-insulator boundary can be defined as the trajectory of $h[\vec{s}]=0$. The distance from the criticality is determined by $(h,\delta r)$ with $\delta r[\vec{s}] \equiv r[\vec{s}]-r_{\rm cr}(u[\vec{s}_{\rm cr}{}^*[\vec{s}]])$, where $\vec{s}_{\rm cr}{}^*[\vec{s}]$ is such $\vec{s}_{\rm cr}$ point as gives a minimum value of $r[\vec{s}]-r(u[\vec{s}_{\rm cr}])$ if $\vec{s}_{\rm cr}$ is not restricted to one point in the $\vec{s}$ space. This implies that for instance, when $\vec{s}$ has three components, there appears a first-order transition surface terminating at a critical end curve. This is consistent with the recently proposed $(U,t',T)$ phase diagram of the two-dimensional square-lattice Hubbard model~\cite{CPM} where $t'$ is the second-neighbor transfer. Therefore, we can fix the transfer parameters and consider changes in $U$ and $T$. We can also fix other short-range Coulomb interaction parameters.

In addition to $U$ and $T$, the chemical potential $\mu$ plays a crucial role in Mott transition, since it controls the electron filling $n$. Quantum Mott transition should take place at a partial but integer filling, and hence there are two MIT boundary values of $\mu$ which correspond to the edges of the Mott gap. This boundary gives a curve in the $(\mu,U)$ space. At finite temperatures, a choice of $n$ uniquely determines $\mu$. Then, the Mott transition extends to small electron and hole doping regions at $T>0$, when it survives as a thermal phase transition. 

When the quantum Mott transition in the $(U,\mu)$ space has a second-order character, then critical fluctuations may immediately smear out the thermal transition in the space of $(U,\mu,T)$. On the contrary, when the quantum Mott transition has a first-order character, then it is robust against the thermal fluctuations and produces a first-order Mott transition surface terminating at a finite-temperature critical end curve. This is because in the ground states on both sides of the first-order transition, there exist neither critical fluctuations nor strong classical fluctuations characterized by power-law or exponential temperature dependence of the correlation length. Although such strong thermal fluctuations yield an abrupt collapse of long-range order containing a finite order parameter amplitude, these fluctuations appear above the quantum second-order transition in the one-dimensional Ising model and in the two-dimensional Heisenberg model~\cite{ChakravartyHalperinNelson89}, but not at the first-order transition in the present case.

Concludingly, the finite-temperature inherent Mott transition is schematically described by a phase diagram shown in Fig.~\ref{fig:phasediagram2}. Here, the first-order transition surface corresponds to a trajectory of $h[U,\mu,T]=0$, while the critical end curve is further specified by $r[U,\mu,T]=r_{\rm cr}(u[U,\mu,T])$. Similar first-order Mott transition surface terminating at a finite-temperature critical end curve has also been discussed in infinite dimensions~\cite{KotliarMurthyRozenberg02}, although we will find novel crucial aspects of critical thermodynamic properties in the rest of this section. 

To recognize the shape of the Mott transition boundary and to discuss critical properties in terms of $\vec{s}$, it is important to note that $h[\vec{s}]$, $r[\vec{s}]$ and $u[\vec{s}]$ are all regular functions, which is guaranteed in the GL arguments. Therefore, we can safely expand $(h,r)$ around $(0,r_{\rm cr})$ in terms of $\vec{s}-\vec{s}_{\rm cr}^{*}$. For instance, $\partial h/\partial \mu$ is well defined around the criticality. Then, as depicted in Fig.~\ref{fig:phasediagram2}, there exists a curve (open dotted curve) determined by $h=\partial h/\partial \mu=0$ with $r<r_c$, where an infinitesimal increase of $U$ but not any change of $\mu$ drives the system from metal to Mott insulator. Namely, there occurs a bandwidth-control Mott transition, but not a filling-control with $U$ being fixed. The shape of the Mott transition surface at finite temperatures is characterized by $dU/d\mu=-(\partial h/\partial \mu)/(\partial h/\partial U)$, which vanishes toward the curve. This means that at least at $T>0$, the surface is free from singularity and particularly rounded around the curve~\cite{note_phasediagram}. Only at $T=0$, there remains a possibility of a nonsingular behavior in the shape of the Mott phase boundary surface with respect to $\mu$, reflecting the Fermi distribution function which appears in expressions for $h$, $r$ and $u$. 

In most of the cases including $d=\infty$~\cite{RMP_DMFT} and $d=2$~\cite{CPM}, the first-order Mott transition surface is not parallel to the $T$ axis in the $(U,\mu,T)$ space and a metallic state appears at the lower temperatures. This is because stronger low-energy quasiparticle excitations in the metallic side contribute to a smaller entropy than in the insulating side. As fluctuations of other degrees of freedom like spins than charges develop, the entropy difference between the metal and the insulator decreases and hence the surface becomes more vertical.

More generally, locations of $h=0$ and $r=r_{\rm cr}(u)$ in the $\vec{s}$ space are nontrivial. For instance, they are not necessarily perpendicular to each other even around the criticality, in contrast to the usual Ising model under the magnetic field. Namely, it may be difficult to find the two routes of $h=0$ and $r=r_{\rm cr}(u)$. This may complicate detailed experimental scaling analyses, as in liquid-gas phase transition.

\subsection{First derivatives of free energy}
\label{subsec:Mott_liquid-gas:first-derivatives}

Here, we reveal singular behaviors of the double occupancy $D$, the electron filling $n$ and the entropy per site $S$. Detailed expressions for general first derivatives of the free energy with respect to $\vec{s}$ are presented in Appendix~\ref{app:derivative:first}. In the vicinity of the criticality, dominant singularities of $D$, $n$ and $S$ are determined by a thermal average of $\varphi$ ($\langle\varphi\rangle$) from
\begin{subequations}
  \begin{eqnarray}
    D[\vec{s}] &\approx& d[\vec{s}_{\rm cr}{}^*]
    -\frac{\partial h}{\partial U}[\vec{s}]\langle\varphi\rangle
    \label{eq:d_exp}\\
    n[\vec{s}] &\approx& n[\vec{s}_{\rm cr}{}^*]
    +\frac{\partial h}{\partial\mu}[\vec{s}]\langle\varphi\rangle
    \label{eq:n_exp}\\
    S[\vec{s}] &\approx& S[\vec{s}_{\rm cr}{}^*]
    +\frac{\partial h}{\partial T}[\vec{s}]\langle\varphi\rangle.
  \label{eq:S_exp}
  \end{eqnarray}
  \label{eq:d,n,S_exp}
\end{subequations}

It is remarkable that as elucidated in Eq.~(\ref{eq:d_exp}), the relevant scalar field $\varphi$ represents a shift of the double occupancy $D$ from the critical value at least around the criticality, since $\partial h/\partial U$ generally does not vanish. Similarly, Eqs.~(\ref{eq:n_exp}) and (\ref{eq:S_exp}) suggest that the field $\varphi$ can also represents the shift of the electron filling $n$ or the entropy $S$ from the critical value only in the case of $\partial h/\partial \mu\ne0$ or $\partial h/\partial T\ne0$, although in the $(U,\mu,T)$ space, there appears a curve of $\partial h/\partial \mu = 0$.

Except the special curve determined by $h=\partial h/\partial \mu=0$ with $r<r_c$ (open dotted curve in Fig.~\ref{fig:phasediagram2}, the double occupancy $D$ and the electron filling $n$ show the same discontinuities as $\langle\varphi\rangle$ at the first-order Mott transition boundary $h=0$ with $r<r_c$ (shaded surface in Fig.~\ref{fig:phasediagram2}). Namely, a first-order Mott transition takes place whether it is bandwidth-control or filling-control. The first-order filling-control transition is nothing but a phase separation. The discontinuities again follow the scaling relations given by Eqs.~(\ref{eq:beta}) and (\ref{eq:delta}). Such discontinuity of $D$ was obtained in previous works~\cite{Castellani79,RMP_DMFT,CPM,RozenbergChitraKotliar99}, and that of $n$ was also obtained~\cite{RMP_DMFT,KotliarMurthyRozenberg02,OnodaImada_jmmm}.

Toward the curve $h=\partial h/\partial \mu=0$, the strength of the singular component $\partial h/\partial \mu$ vanishes. Although the strength may have a linear dependence on a distance from the curve, it depends on details of a structure of the metal-insulator phase boundary.

At the special curve $h=\partial h/\partial \mu=0$, $\mu$ is eventually decoupled from the field $h$ conjugate to $\varphi$ and hence $n$ is decoupled from $\langle\varphi\rangle$ up to a linear order. This means that only a first-order bandwidth-control transition takes place there but not a filling-control. Then, $n$ is continuous and a phase separation does not occur. Actually, an infinitesimal deviation of $\mu$ around the special curve always gives a metallic state on the both sides.

Lastly, we discuss the entropy $S$. Since the Mott transition surface has a slope against a temperature variation as already noted, $\partial h/\partial T$ does not necessarily vanish on the surface even at the critical end curve, and hence there appears a discontinuity of $S$ which scales as $\langle\varphi\rangle$. This leads to a finite latent heat with varying the temperature across the first-order Mott transition surface, as in liquid-gas phase transitions.

\subsection{Second derivatives of free energy}
\label{subsec:Mott_liquid-gas:second-derivatives}

Here we reveal critical properties of susceptibilities $\chi_{DD}$, $\chi_{Dn}$ and $\chi_{nn}$, specific heat $C_{U,\mu}=-T\left(\frac{\partial^2 F}{\partial T^2}\right)_{U,\mu}$ at constant $U$ and $\mu$, and latent heats $L_\mu=-T\left(\frac{\partial^2 F}{\partial T \partial U}\right)_\mu=T\left(\frac{\partial S}{\partial U}\right)_{\mu,T}$ and $L_U=-T(\frac{\partial^2 F}{\partial T \partial \mu})_U=T\left(\frac{\partial S}{\partial \mu}\right)_{U,T}$. Expressions for general second derivatives of the free energy with respect to $\vec{s}$ are presented in Appendix~\ref{app:derivative:second}. 

First, we discuss susceptibilities. Their singularities in the vicinity of the criticality are given by
\begin{subequations}
  \begin{eqnarray}
  \chi_{DD}
  &\approx&(\frac{\partial h}{\partial U})^2\chi
  +(\frac{\partial r}{\partial U})^2\chi_4
  +\frac{\partial^2 h}{\partial U^2}\langle\varphi\rangle,
  \label{eq:chi_DD_exp}
  \\
  \chi_{Dn}
  &\approx&\frac{\partial h}{\partial \mu} \frac{\partial h}{\partial U}\chi
  +\frac{\partial r}{\partial U}\frac{\partial r}{\partial \mu}\chi_4
  +\frac{\partial^2 h}{\partial U \partial \mu}\langle\varphi\rangle,
  \label{eq:chi_Dn_exp}
  \\
  \chi_{nn}
  &\approx&(\frac{\partial h}{\partial \mu})^2\chi
  +(\frac{\partial r}{\partial \mu})^2\chi_4
  +\frac{\partial^2 h}{\partial \mu^2}\langle\varphi\rangle,
  \label{eq:chi_nn_exp}
  \end{eqnarray}
\end{subequations}
with $\chi_4$ being given in Eq.~(\ref{eq:chi_4}). Therefore, the doublon susceptibility $\chi_{DD}$ diverges toward the criticality as obtained for the Hubbard model~\cite{CPM,RozenbergChitraKotliar99}, and the critical exponents are the same as those for $\chi$ given in Eqs.~(\ref{eq:gamma}) and (\ref{eq:delta_chi}). Similarly, the charge compressibility $\chi_{nn}$ and $\chi_{Dn}$ also diverge at the critical end curve (thick solid curve in Fig.~\ref{fig:phasediagram2}) except the special critical end point $(\tilde{U}_{\rm cr},\tilde{\mu}_{\rm cr},\tilde{T}_{\rm cr})$ determined by $\delta r=h=\partial h/\partial \mu=0$ (open circle in Fig.~\ref{fig:phasediagram2}). Similar compressibility divergence was obtained in infinite dimensions~\cite{KotliarMurthyRozenberg02}. At $(\tilde{U}_{\rm cr},\tilde{\mu}_{\rm cr},\tilde{T}_{\rm cr})$, while $\chi_{DD}$ diverges as already noted, $\chi_{Dn}$ and $\chi_{nn}$ become regular, in contrast to the other points in the critical end curve. This is because $\partial h/\partial \mu$ and thus the second terms of Eqs.~(\ref{eq:chi_Dn_exp}) and (\ref{eq:chi_nn_exp}) vanish there.

Next, we discuss a scaling behavior of $C_{U,\mu}$, $L_\mu$ and $L_U$. These quantities can also be expressed as
\begin{subequations}
  \begin{eqnarray}
  C_{U,\mu}
  &\approx& (\frac{\partial h}{\partial T})^2\chi
  +(\frac{\partial r}{\partial T})^2\chi_4
  +\frac{\partial^2 h}{\partial T^2}\langle\varphi\rangle
  \label{eq:C}
  \\
  L_\mu
  &\approx& \frac{\partial h}{\partial T} \frac{\partial h}{\partial U}\chi
  +\frac{\partial r}{\partial T}\frac{\partial r}{\partial U}\chi_4
  +\frac{\partial^2 h}{\partial U \partial T}\langle\varphi\rangle
  \label{eq:L_mu}
  \\
  L_U
  &\approx& \frac{\partial h}{\partial T} \frac{\partial h}{\partial \mu}\chi
  +\frac{\partial r}{\partial T}\frac{\partial r}{\partial \mu}\chi_4
  +\frac{\partial^2 h}{\partial \mu \partial T}\langle\varphi\rangle.
  \label{eq:L_U}
  \end{eqnarray}
\end{subequations}
When $\partial h/\partial T$ does not vanish due to a finite slope of the Mott transition surface with respect to temperature, singularity of $C_{U,\mu}$ is dominated by the first term in Eq.~(\ref{eq:C}) and hence the critical properties are determined by those of $\chi$ given in Eq.~(\ref{eq:scaling}), as in liquid-gas phase transitions~\cite{Ma}. Namely,
\begin{subequations}
\begin{eqnarray}
  C_{U,\mu}|_{r,h=0} &\propto& |r-r_{\rm cr}|^{-\gamma} \propto |T-T_{\rm cr}|^{-\gamma}
  \label{eq:C_Umu_gamma}\\
  C_{U,\mu}|_{r=r_{\rm cr},h} &\propto& |h|^{-1+1/\delta} \propto |T-T_{\rm cr}|^{-1+1/\delta}.
  \label{eq:C_Umu_delta}
\end{eqnarray}
\label{eq:C_Umu1}
\end{subequations}
Otherwize, it reduces to the conventional scaling relations for the specific heat with $h$ being fixed constant;
\begin{subequations}
\begin{eqnarray}
  C_h|_{r,h=0} &\propto& |r-r_{\rm cr}|^{-\alpha} \propto |T-T_{\rm cr}|^{-\alpha}
  \label{eq:C_Umu_alpha}\\
  C_h|_{r=r_{\rm cr},h} &\propto& |h|^{-\alpha/\beta\delta} \propto |T-T_{\rm cr}|^{-\alpha/\beta\delta}.
  \label{eq:C_Umu_alpha_gamma}
\end{eqnarray}
\label{eq:C_Umu0}
\end{subequations}
The same scaling properties applies to $L_{\mu}$ and $L_U$, except at $(\tilde{U}_{\rm cr},\tilde{\mu}_{\rm cr},\tilde{T}_{\rm cr})$ where $L_U$ becomes regular due to vanishing singular components.

It is remarkable that the obtained divergence of specific heat $C_h$ at the critical end curve for $d\le d_c=4$ contrasts to MF arguments such as the DMFT~\cite{RMP_DMFT} and the Brinkman-Rice~\cite{BrinkmanRice70} or slave-boson MF~\cite{KotliarRuckenstein86} theory. In the MF universality which is indeed realized in $d>d_c$, the specific heat and thus the quasiparticle effective mass do not diverge for $T>0$ even at criticality. This reflects that the inverse of the quasiparticle renormalization factor,
\begin{equation}
  1/z({\bf k})
  =1-\frac{\partial {\rm Re}\Sigma}{\partial \omega}(\omega=0,{\bf k}),
\label{eq:z_k}
\end{equation}
remains finite at the finite-temperature criticality. As we have obtained above, $C_h$ diverges at the finite-temperature critical end curve for $2\le d\le d_c$. The diverging quantity is, however, not $1/z({\bf k})$ but the inverse of the suppression factor of the Fermi velocity due to spatial fluctuations,
\begin{equation}
  v^{\rm b}({\bf k})/v^{\rm r}_k({\bf k})= \left[1+\frac{\partial {\rm Re}\Sigma}{\partial k_\bot}(\omega=0,{\bf k})/v({\bf k})\right]^{-1}
\label{eq:v_k}
\end{equation}
with $k_\perp$ being the component of ${\bf k}$ perpendicular to the Fermi surface. For $d=\infty$, $v^{\rm b}({\bf k})/v^{\rm r}_k({\bf k})=1$, since there exists no ${\bf k}$ dependence of the self-energy part; a regular behavior of $C_h$ in the case of $d=\infty$ is immediately understood as a typical MF behavior. We can obtain a further insight into the critical properties of the Fermi velocity $v^{\rm r}({\bf k})=z({\bf k})v^{\rm r}_k({\bf k})$ in finite dimensions. The entropy $S$ of the quasiparticle system is given by the following ${\bf k}$ integral on the Fermi surface $S_{\rm F}$,
\begin{equation}
S \propto T\int_{S_{\rm F}}\!\frac{da_{\bf k}}{(2\pi)^{d-1}}\frac{1}{z({\bf k})v^{\rm r}_k({\bf k})},
\label{eq:entropy_quasiparticle}
\end{equation}
where $a_{\bf k}$ represents a $(d-1)$-dimensional infinitesimal surface element around ${\bf k}$. We assume that both $v^{\rm r}_k({\bf k})$ and $v^{\rm r}_k({\bf k})$ vanish most rapidly toward a certain Fermi momentum ${\bf k}_F^0$ as $\delta k_\perp^\varpi$, where $\delta k_\perp$ is the amplitude of the tangential component of ${\bf k}-{\bf k}_F^0$ to the Fermi surface. Since from Eq.~(\ref{eq:C_Umu0}) a scaling dimension of $S$ is $(\alpha-1)/\nu=-d+1/\nu$ and $z({\bf k})$ is finite, then $v({\bf k})$ should have a scaling dimension of $1/\nu+\varpi-1$. Namely,
\begin{eqnarray}
v^{\rm r}({\bf k}) & = & z({\bf k})v^{\rm r}_k({\bf k}) \propto \xi^{-\zeta/\nu} \propto |T-T_{\rm cr}^*|^{\zeta},
\label{eq:v_scaling}\\
\zeta &=&1-\nu(1-\varpi).
\label{eq:upsilon}
\end{eqnarray}
Particularly, in the case of a linear dispersion relation $\varpi=0$ on the whole Fermi surface, toward the criticality along the metal-insulator boundary $h=0$, $v^{\rm r}({\bf k})$ vanishes as $1/\ln|T-T_{\rm cr}^*|$ in two dimensions and as $|T-T_{\rm cr}^*|^{0.36}$ in three dimensions. If the bare Fermi velocity vanishes as $\delta k_\parallel$ as in a van-Hove singularity, namely, $\varpi=1$, then $v^{\rm r}({\bf k})$ vanishes as $\delta k_\parallel |T-T_{\rm cr}^*|$ in $2 \le d < d_c$. These sharply contrast to the case of $d>d_c$ where $v^{\rm r}/v^{\rm b}$ does not vanish.

\subsection{Discussion}
\label{subsec:Mott_liquid-gas:discussion}

As already noted~\cite{GL_DMFT,OnodaNagaosa03}, the same singular behaviors as $\langle\varphi\rangle$ or $\chi$ can be observed in many other physical quantities for transport, magnetic and optical properties. However, in spite of intensive and extensive experimental investigations on the Mott transition systems like V$_{2-x}$(Cr/Ti)$_x$O$_{3-y}$, NiS$_{2-x}$Se$_x$, $\kappa$-(BEDT-TTF)$_2$X, etc, there are only a few systematic experimental studies of the thermal criticality of Mott transition~\cite{Science_Limelette03}. Here, we discuss possibilities of experimentally verifying the critical behaviors from magnetic, transport, optical and spectral measurements.

First, we discuss the dc longitudinal resistivity $\rho$. Around the Mott criticality, it can be expanded in $\langle\varphi\rangle$. Then, $\rho$ and $\partial \rho/\partial \vec{s}$ follow the same scaling relations as $\langle\varphi\rangle$ and $\chi$, respectively, in other words, $D$ and $\chi_{DD}$. This is consistent with recent experiments on V$_{2-x}$Cr$_x$O$_3$ by Limelette {\it et al.}~\cite{Science_Limelette03}. Similar discontinuity of $\rho$ was also found in NiS$_2$ though scaling analysis has not been performed~\cite{Anzai86}.

We proceed to magnetic properties. In the case of single-band Hubbard model, the double occupancy $D$ and the electron filling $n$ directly determine the local moment $\langle {S_{\rm loc}^z}^2 \rangle^{1/2} = (1/2)\sqrt{n-2D}$. Therefore, if spin correlation length is so short that the uniform spin susceptibility $\chi_{SS}$ follows the Curie-Weiss law, then the Curie constant should be proportional to $(n-2D)/4$ and hence directly show the singularities of $D$ and $n$~\cite{MenthRemeika70}. Namely, the Curie constant exhibits a discontinuity at the first-order transition, whether we approach the first-order Mott transition surface in Fig.~\ref{fig:phasediagram2} by decreasing the carrier doping concentration or changing $U/W$. More generally, $\chi_{SS}$ can be expanded in terms of $\langle \varphi \rangle$. It discontinuously jumps at the first-order transition and $\partial\chi_{SS} /\partial T$ diverges at the criticality. The discontinuity was experimentally observed, though scaling analysis has not been performed. The critical exponents are given by those for $D$ and $\chi_{DD}$. The same critical properties as $\chi_{SS}$ apply to the Knight shift and the spin-lattice relaxation rate.

In order to probe the detailed nature at the first-order Mott transition, NMR measurements are useful when there exists a sufficiently strong coupling of the nuclear spins to the electron near the chemical potential~\cite{Slichter}. By observing a splitting of the NMR/NQR spectral peak into two, one can detect an emergence of a macroscopic inhomogeneity and coexistence of the metallic and the insulating states at the first-order Mott transition boundary, which have different values of $D$, and even $n$ when the filling-control Mott transition also occurs. It is more likely that in the time scale of the NMR, macroscopic coexistence disappears and the globally stable state is realized. In this case, the peak discontinuously jumps. This may be experimentally verified in the future. So far, NMR measurements in V$_{2-x}$(Cr/Ti)$_x$O$_{3-y}$~\cite{GossardMcWhanRemeika70} were done mainly at low temperatures, but neither around the Mott critical end point nor along the first-order Mott transition boundary. Alloying for a carrier doping into Mott insulator induces disorder effects. Nevertheless, around the thermal Mott criticality, the above critical properties should dominate over the disorder effects.

The divergence of the charge compressibility should be observed as the dip of the sound velocity via ultrasonic~\cite{Fournier} and Brillouin scattering measurements. This reflects that the electronic contribution to the sound velocity, which is inversely proportional to $\chi_{nn}$, vanishes at the critical end curve except at $(\tilde{U}_{\rm cr},\tilde{\mu}_{\rm cr},\tilde{T}_{\rm cr})$. Fournier {\it et al}. has observed the critical enhancement of $\chi_{nn}$~\cite{Fournier} by the ultrasonic measurements, though scaling analyses have not been performed.

Now we discuss spectral properties such as spectral weight of coherent one-electron excitations and Drude weight. As in $d=\infty$, $\langle\varphi\rangle$ also measures the coherent spectral weight $w$ observed in photoemission; $w=w_{\rm cr}+w'\langle\varphi\rangle$ around the criticality where $w_{\rm cr}$ is a value of $w$ at a critical point and does not vanish in general. This property of $w$ can also be detected through optical measurements. In many cases, the optical conductivity $\sigma(\omega)$ nearly splits into low-energy coherent and high-energy incoherent spectral weights. While the latter part appears above the Hubbard band gap energy, the former has a total weight proportional to $w^2$. In particular, the Drude weight for the lowest-energy excitations has also the same critical properties as $D$.

In the rest of this section, we discuss implications of the present theory. The distinction of metal and insulator is strict at $T=0$ where the dc conductivity can vanish in the insulator. On the contrary, at $T>0$, a small but nonzero conductivity appears due to thermal activation. This makes it difficult to identify the insulator at finite temperature. Nevertheless, it is meaningful to categorize the Mott insulator as a distinct state from metal as in the present arguments. This is possible, because around integer but partial electron fillings, two distinct states, a globally stable and a semi-stable ones, appear near the first-order transition terminating at criticality. We have shown that these states, namely, a strongly correlated Mott insulator and a less correlated metal, correspond to the oppositely polarized configurations of the Ising model, as discussed by Kotliar {\it et al.} for $d=\infty$~\cite{GL_DMFT}.

The universal properties obtained for $D$, $n$, $\chi_{DD}$, $\chi_{Dn}$ and $\chi_{nn}$ implies that if one could construct a GL free energy expansion in terms of two fields $\phi_D$ and $\phi_n$ representing $D$ and $n$ by appropriately treating the local constraints in a slave-particle representation, then the coefficients to $\phi_D^2$, $\phi_D\phi_n$ and $\phi_n^2$ should vanish similarly at the criticality, except the special bandwidth-control critical point $(\tilde{U}_{\rm cr},\tilde{\mu}_{\rm cr},\tilde{T}_{\rm cr})$ where the field $\phi_n$ becomes massive. Furthermore, the present theory poses a severe constraint on unified analyses of bandwidth-control and filling-control Mott transitions; if the bandwidth-control Mott transition is of the first order at the open dotted curve $h=\partial h/\partial \mu=0$ in Fig.~\ref{fig:phasediagram2}, the Mott transition at the shaded surface are also of the first-order, whether the control parameter is $D$ or $n$. This reflects that two fields $\phi_D$ and $\phi_n$ appearing in such analyses in terms of the slave-particle representation have the same origin of the charge degrees of freedom and must be described by a scalar field $\varphi$. Such free energy functional in terms of both $\phi_D$ and $\phi_n$ is still not easy to handle due to difficulty in treating the local constraints.

\section{Interplay between the Inherent Mott Transition and Other Symmetry-Breaking Transitions}
\label{sec:interplay}

So far, we have considered a prototype of Mott transition occurring without any spontaneous symmetry breaking, which has a phase diagram shown in Fig.~\ref{fig:phasediagram} (a). In this section, we discuss its interplay with other spontaneously symmetry-breaking transitions. 

\subsection{Mean-field analyses}
\label{subsec:interplay:MF}

We concentrate on the interplay between the scalar field $\varphi$ and the $O(m)$-symmetric vector field $\vec{\phi}$ representing the charge and other degrees of freedom, respectively. For instance, $m=1$, $2$ and $3$ describe interplay with charge ordering, superconductivity and magnetism, respectively. Then, according to Eq.~(\ref{eq:F_GL}), our GL free energy can be written in the form
\begin{eqnarray}
  \lefteqn{\frac{F_{\rm GL}}{T} = \int\!d{\bf x}\Big[-h\varphi({\bf x})
      +\frac{1}{2}\left(r_\varphi \varphi({\bf x})^2
      +({\bf \nabla}\varphi({\bf x}))^2\right)}
      \nonumber\\
      &&+\frac{1}{2}\left(r_\phi\vec{\phi}({\bf x})^2+({\bf \nabla}\vec{\phi}({\bf x}))^2\right)
      +\frac{g}{2}\varphi({\bf x})\vec{\phi}({\bf x})^2
      \nonumber\\
      &&+\frac{u_{\varphi\varphi}}{4}\varphi({\bf x})^4
      +\frac{u_{\varphi\phi}}{2}\varphi({\bf x})^2\vec{\phi}({\bf x})^2
      +\frac{u_{\phi\phi}}{4}(\vec{\phi}({\bf x})^2)^2\Big].
  \label{eq:F_GL_interplay}
\end{eqnarray}
$r_\varphi$ and $r_\phi$ are the second-order coupling constants for $\varphi$ and $\phi$, respectively. $g_\varphi$ and $g_\phi$ denote the third-order coupling constants for $\varphi$ and the mixing of $\varphi$ and $\phi$, respectively. $u_{\varphi\varphi}$ and $u_{\phi\phi}$ represent the fourth-order coupling constants of $\varphi$ and $\phi$, respectively, while $u_{\varphi\phi}$ that of their mixing. Within the GL theory, phase diagrams of the above free energy are given by Fig.~\ref{fig:phasediagram} (b-d).

In the case of Fig.~\ref{fig:phasediagram} (b), the inherent Mott transition, which occurs without any spontaneous symmetry breaking, is removed by emergence of the antiferromagnetic insulating phase, and only a Mott metal-insulator crossover remains at the high-temperature phase. This occurs when $r_\varphi-r_\phi u_{\varphi\phi}/u_{\phi\phi}>0$ holds at $h=-gr_\phi/(2u_{\phi\phi})$ in the ordered phase $r_\phi<0$. Physically, this corresponds to the case where a partially filled band disappears in the reduced Brillouin zone of this antiferromagnetic phase and hence it smears out a phase transition only associated with the charge degrees of freedom. Then, this phase transition is an Slater-type Mott transition. The phase diagram appears in $d>2$ when a geometrical frustration is so weak that it becomes much larger than the Mott temperature scale determined by $r_\varphi=0$. 

In the case of Fig.~\ref{fig:phasediagram} (c1) or (c2), for instance, antiferromagnetic phase transitions take place on the both sides of Mott transition. Antiferromagnetic metal and antiferromagnetic insulator are separated by the first-order Mott transition at low temperatures. This occurs when $r_\phi \pm g\sqrt{-r_\varphi/u_{\varphi\varphi}}-r_\varphi u_{\varphi\phi}/u_{\varphi\varphi}$ becomes negative with decreasing temperature below the Mott critical temperature $T_{\rm cr}$ along the first-order Mott transition boundary. Discontinuity of $\langle\varphi\rangle$ at the first-order Mott transition boundary introduces a jump $2g\sqrt{-r_\varphi/u_{\varphi\varphi}}$ in the mass term of $\phi$, producing a difference in transition temperatures on the both sides of the first-order Mott transition. The phase diagram (c1) has been observed in NiS$_{2-x}$Se$_x$ and $R$NiO$_3$ for interplay with AF, where antiferromagnetic metal and insulator are separated by the first-order Mott transition at low temperatures. Physically, this corresponds to the case where effects of the geometrical frustration increase from the case of (b) and the Mott criticality appears above the Neel temperature. The phase diagrams in (c) are also relevant to the case where there appears the interplay of Mott criticality with multicritical phenomenon of SC and AF~\cite{OnodaNagaosa03} as in $\kappa$-(BEDT-TTF)$_2$X. 

In the case of Fig.~\ref{fig:phasediagram} (d1) or (d2), an antiferromagnetic phase or a superconducting phase appears at low temperatures in the insulating or the metallic side of the first-order Mott transition. This occurs when only $r_\phi - |g|\sqrt{-r_\varphi/u_{\varphi\varphi}}-r_\varphi u_{\varphi\phi}/u_{\varphi\varphi}$ but not $r_\phi + |g|\sqrt{-r_\varphi/u_{\varphi\varphi}}-r_\varphi u_{\varphi\phi}/u_{\varphi\varphi}$ turns to be negative with decreasing temperature below $T_{\rm cr}$ along the first-order Mott transition boundary. In the issue of interplay with antiferromagnetism, this corresponds to systems with geometrical frustration further amplified from the case of (c), smearing out the antiferromagnetic metal. The phase diagram (d1) has been observed in such systems as V$_{2-x}$Cr$_x$O$_3$~\cite{Mott,RMP_ImadaFujimoriTokura} and $\kappa$-(BEDT-TTF)$_2$X under a magnetic field~\cite{Kagawa_unpublished}. The phase diagram (d2) is relevant to $\kappa$-(BEDT-TTF)$_2$Cu$_2$(CN)$_3$~\cite{ShimizuMiyagawaKanodaMaesatoSaito03} for interplay with SC and $R_2$Mo$_2$O$_7$~\cite{Tokura_MIT_pyrochlore} for that with ferromagnetism.

Within phenomenological GL arguments, it is also possible that a first-order transition and its critical end point emerge in the ordered phase, though it is not shown in Fig.~\ref{fig:phasediagram}. This occurs when $r_\varphi-r_\phi u_{\varphi\phi}/u_{\phi\phi}$ becomes negative along the curve of $h=-gr_\phi/u_{\phi\phi}$ below a certain temperature in the symmetry-broken phase $r_\phi<0$. To our knowledge, such phase diagram has not been experimentally observed in correlated electron systems. It may occur in {\it charge-ordered spin-disordered phase of quarter-filled systems}, where once the staggered charge ordering appears then the system effectively corresponds to a half-filled one with the spin symmetry being unbroken. This phenomenon can also be understood by the present framework from the following reason: $\varphi$ and $\phi$ represent a uniform and a staggered charge densities, respectively. While the Hamiltonian does not generally contain a symmetry under $\varphi\to-\varphi$, it is invariant under $\phi\to-\phi$. This leads to the free energy of Eq.~(\ref{eq:F_GL_interplay}).

\subsection{Fluctuation-induced first-order phase transitions}
\label{subsec:interplay:1st-order}

If the critical regions of two degrees of freedom overlap appreciably in the phase diagrams, then interplay between these critical fluctuations often alter structures of the phase diagrams. Below, taking into account critical fluctuations by means of one-loop renormalization group (RG) analyses, we show that fluctuation-induced first-order transitions as multicritical phenomena may occur in the vicinity of the meeting points of the transitions.

We introduce a momentum cutoff $\Lambda=e^\lambda$, and define dimensionless coupling constants $\tilde{h}\equiv \Lambda^{-1-d/2}h/\sqrt{I_d}$, $\tilde{r}_i\equiv\Lambda^{-2}I_dr_i$, $\tilde{g}\equiv\Lambda^{-3+d/2}\sqrt{I_d}g$ and $\tilde{u}_{ij}\equiv\Lambda^{-4+d}I_d u_{ij}$ with $I_d=2[(4\pi)^{d/2}\Gamma(d/2)]^{-1}$. Then, we perform renormalization group (RG) analyses. Up to one-loop order, RG equations are obtained as
\begin{subequations}
  \begin{eqnarray}
    \frac{d\tilde{g}}{d\lambda}
    &=&[-\varepsilon_3+(m+2)\tilde{u}_{\varphi\phi}+(m+8)\tilde{u}_{\phi\phi}
    +\tilde{u}_{\varphi\phi}^2/(6\tilde{u}_{\varphi\varphi})
    \nonumber\\
    &&-\tilde{g}^2(1+m\tilde{u}_{\varphi\phi}/(6\tilde{u}_{\varphi\varphi}))] \tilde{g}
    \label{eq:RG:g}\\
    \frac{d\tilde{u}_{\varphi\varphi}}{d\lambda}
    &=&-\varepsilon \tilde{u}_{\varphi\varphi}
    +9\tilde{u}_{\varphi\varphi}^2+m\tilde{u}_{\varphi\phi}^2
    +\frac{m}{2}\tilde{g}^4
    \label{eq:RG:u_varphivarphi}\\
    \frac{d\tilde{u}_{\phi\phi}}{d\lambda}
    &=&-\varepsilon\tilde{u}_{\phi\phi}+(m+8)\tilde{u}_{\phi\phi}^2
    +\tilde{u}_{\varphi\phi}^2+\tilde{g}^4
    \label{eq:RG:u_phiphi}\\
    \frac{d\tilde{u}_{\varphi\phi}}{d\lambda}
    &=&[-\varepsilon+3\tilde{u}_{\varphi\varphi}+(m+2)\tilde{u}_{\phi\phi}
      +4\tilde{u}_{\varphi\phi}]\tilde{u}_{\varphi\phi}+\tilde{g}^4,
    \nonumber\\
    \label{eq:RG:u_varphiphi}
  \end{eqnarray}
  \label{eq:RG}
\end{subequations}
with $\varepsilon_3=3-\frac{d}{2}>0$ and $\varepsilon=4-d>0$. 

\begin{figure}[htb]
\begin{center}
\includegraphics[width=7.8cm]{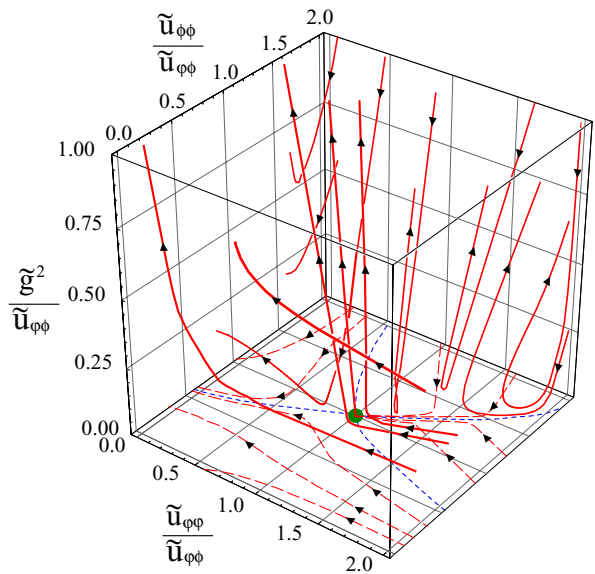}
\end{center}
\caption{(color online) RG flow diagram in the space of $(\tilde{u}_{\varphi\varphi}/\tilde{u}_{\varphi\phi},\tilde{u}_{\phi\phi}/\tilde{u}_{\varphi\phi},\tilde{g}^2/\tilde{u}_{\varphi\phi})$ for $m=3$ and $d=3$. Dashed curves represent RG flows in the plane of $\tilde{g}^2/\tilde{u}_{\varphi\phi}=0$. They are separated by the two dotted curves. Solid curves represent RG flows for $\tilde{g}^2/\tilde{u}_{\varphi\phi}>0$.}
\label{fig:flow}
\end{figure}

There exist three fixed points of $(\tilde{g},\tilde{u}_{\varphi\varphi},\tilde{u}_{\varphi\phi},\tilde{u}_{\phi\phi})$; a decoupled Ising-MF fixed point $(0,\varepsilon/9,0,0)$, a decoupled Ising-$O(m)$ fixed point $(0,\varepsilon/9,\varepsilon/(m+8),0,0)$ and the $O(m+1)$-symmetric fixed point $(0,\varepsilon/(m+9),\varepsilon/(m+9),\varepsilon/(m+9))$. They are all unstable to an infinitesimal value of $\tilde{g}$. The two decoupled fixed points are characterized by a simultaneous emergence of the inherent Mott criticality for $\varphi$ and $O(m)$-symmetric one for $\phi$ which are decoupled from each other. It is also unstable to $\tilde{u}_{\varphi\phi},\tilde{u}_{\phi\phi}\ne0$. The $O(m+1)$-symmetric fixed point leads to a bicritical behavior and is only partly stable within the space of $(\tilde{u}_{\varphi\varphi},\tilde{u}_{\varphi\phi},\tilde{u}_{\phi\phi})$. We note that up to the one-loop order, for $m=2$ or $3$, there does not exist a biconical fixed point which usually leads to a tetracritical behavior. In Fig.~\ref{fig:flow}, we show RG flow trajectories of the parameters $(\tilde{u}_{\varphi\varphi}/\tilde{u}_{\varphi\phi},\tilde{u}_{\phi\phi}/\tilde{u}_{\varphi\phi},\tilde{g}/\tilde{u}_{\varphi\phi})$ in the case of $m=3$, $d=3$ and $\tilde{u}_{\varphi\phi}>0$.

In the plane of $\tilde{g}^2/\tilde{u}_{\varphi\phi}=0$, the region nearly determined by $\tilde{u}_{\varphi\varphi}\tilde{u}_{\phi\phi}/\tilde{u}_{\varphi\phi}^2\gtrsim 1$ shrinks into the $O(m+1)$-symmetric fixed point as RG flows show. The other part of the plane flows toward $\tilde{u}_{\varphi\varphi}/\tilde{u}_{\varphi\phi}\to-\infty$ and/or $\tilde{u}_{\phi\phi}/\tilde{u}_{\phi\varphi}\to-\infty$. Such runaway flows toward negative values of $u_{\varphi\varphi}$ and/or $u_{\phi\phi}$ indicate a fluctuation-induced first-order phase transition~\cite{Amit}. Usually, a tricritical behavior appears in the case of runaway flows. 

A finite third-order coupling $\tilde{g}$ renders the interplay between $\varphi$ and $\phi$ more severe, which is also displayed in Fig~\ref{fig:flow}. The flows are largely categorized into three depending on the bare coupling constants:

(I) If the bare coupling constants satisfy $\tilde{u}_{\varphi\varphi}\tilde{u}_{\phi\phi}/\tilde{u}_{\varphi\phi}^2\gg 1$, the parameters $(\tilde{u}_{\varphi\varphi},\tilde{u}_{\varphi\phi},\tilde{u}_{\phi\phi})$ first flow toward the $O(m+1)$-symmetric fixed point and then quickly turn to go to infinity. Then, $\tilde{u}_{\varphi\phi}$ becomes negative and sign changes of $\tilde{u}_{\varphi\varphi}$ and $\tilde{u}_{\phi\phi}$ follow, although it is not shown in the figure. In this case, the system undergoes a fluctuation-induced first-order phase transition to a phase which is likely to have both long-range orders $\langle\varphi\rangle,\langle\phi\rangle\ne0$. However, detailed properties of the ordered phase actually depends on higher-order terms which are not included in Eq.~(\ref{eq:F_GL_interplay}).

(II) If the bare coupling constants satisfy $\tilde{u}_{\varphi\varphi}\tilde{u}_{\phi\phi}/\tilde{u}_{\varphi\phi}^2\ll 1$, the parameters flow toward infinity. In this case, $\tilde{u}_{\varphi\varphi}$ and/or $\tilde{u}_{\phi\phi}$ turn to be negative while $\tilde{u}_{\varphi\phi}$ remains positive. This leads to a fluctuation-induced first-order transition to a phase which is likely to have only an long-range order of either $\varphi$ or $\phi$. 

(III) We consider the cases where the bare coupling constants satisfy $\tilde{u}_{\varphi\varphi}\tilde{u}_{\phi\phi}/\tilde{u}_{\varphi\phi}^2\sim 1$. In this case, the parameters $(\tilde{u}_{\varphi\varphi},\tilde{u}_{\varphi\phi},\tilde{u}_{\phi\phi})$ first approach the $O(m+1)$-symmetric fixed point, and the flow is modified by the fixed point into a nearly vertical one. Then, while the relation $\tilde{u}_{\varphi\varphi}\sim\tilde{u}_{\varphi\phi}\sim\tilde{u}_{\phi\phi}$ is approximately satisfied, $\tilde{g}^2/\tilde{u}_{\varphi\phi}$ rapidly increases. This gives rise to a bicritical scaling behavior away from the transition. Finally, the parameters flow to infinity and the fluctuation-induced first-order phase transition noted in either (I) or (II) appears instead of the bicriticality.

\section{Summary}
\label{sec:Summary}

We have developed Ginzburg-Landau theory of classical Mott transition for the dimensionality $d \le \infty$ by reexamining the framework of the cluster dynamical mean-field theory. Then, we have confirmed that the thermal critical phenomena for the Mott transition occurring without any spontaneous symmetry breaking can be described in terms of a liquid-gas phase transition characterized by the Ising universality class, in accordance with a recent experimental scaling analysis of the resistivity~\cite{Science_Limelette03}. There appears a relevant scalar field $\varphi$ representing the charge degrees of freedom characterized by the double occupancy. This plays a role of ``order parameter'' and describes a bosonic soft mode at Mott criticality. In the space of the local Coulomb repulsion $U$, the chemical potential $\mu$ and the temperature $T$, there appears a first-order Mott transition surface terminating at a critical end curve. Around the criticality, the double occupancy, the electron filling and the entropy have $\varphi$-linear terms. They show discontinuities and diverging responses at the first-order and the second-order Mott transitions, respectively. The divergence of specific heat indicates that of quasiparticle effective mass at the criticality. There exists a curve where the Mott gap closes. At this curve, a filling-control transition does not occur but a bandwidth-control, the $\varphi$-linear term in the electron filling vanishes, and the related singularities disappear. We have also provided principles for experimental tests of the proposed critical scaling behaviors and emergence of a phase separation from transport, magnetic and spectral measurements.

Various experimentally observed phase diagrams have been discussed in terms of the GL theory for interplay between the inherent Mott transition occurring without any symmetry breaking and a symmetry-breaking phase transition such as charge ordering, superconducting and antiferromagnetic transitions. It has also been shown that Mott transition accompanied by a symmetry breaking is described by the symmetry-breaking phase transition under renormalization effects in the proximity of the Mott criticality. In the presence of two critical degrees of freedom, there is a possibility of observing a bicritical phenomena as a crossover. If the renormalization becomes strong, even a fluctuation-induced first-order transition occurs as a multicritical phenomenon. 

Model calculations of the obtained critical properties and an application to quantum Mott transition are left for future studies. Here, we only note that present arguments can not be directly extended for addressing issues of quantum Mott transition and quantum-classical crossover. This is because at $T=0$, susceptibilities $\chi_{DD}$, $\chi_{Dn}$ and $\chi_{nn}$ vanish in the Mott insulating states, in contrast to the case of $T>0$. The present framework is also applicable to other correlation-driven metal-insulator transitions, like those associated with a charge ordering phenomenon.

\begin{acknowledgments}
The author is grateful to N. Nagaosa for stimulating discussion, and F. Mancini, A. Avella and S. Odashima for hospitality during his stay in Salerno. He also thanks K. Kanoda, F. Kagawa, Y. Tokura, Y. Motome, and A. Furusaki for fruitful discussion, and M. Imada for useful comments at an early stage of this work.
\end{acknowledgments}

\begin{appendix}

\section{$U$, $\mu$ and $T$ Derivatives of the Free Energy}
\label{app:derivative}

\subsection{First derivatives}
\label{app:derivative:first}

Here, we give an expression of the first derivative with respect to $s_j$ which is a $j$th component of $\vec{s}$;
\begin{eqnarray}
  &&\frac{1}{N_c}\frac{\partial F}{\partial s_j}[\vec{s}]
  = \frac{1}{N_c}\langle \frac{\partial F_{\rm GL}}{\partial s_j} \rangle
  \nonumber\\
  &=&\frac{1}{N_c}\!\int\!\!d{\bf x}\,\langle
  -\frac{\partial h}{\partial s_j}\varphi({\bf x})
  +\frac{1}{2}\frac{\partial r}{\partial s_j}\varphi({\bf x})^2
  +\frac{1}{4}\frac{\partial u}{\partial s_j}\varphi({\bf x})^4\rangle
  \nonumber\\
  &=& \frac{1}{N_c}\frac{\partial F}{\partial s_j}[\vec{s}_{\rm cr}{}^*]
  +a^{(1)}_{s_j}[\vec{s}]\langle\varphi\rangle
  +a^{(2)}_{s_j}[\vec{s}]\langle\varphi\rangle^2+\cdots
  \label{eq:d}
\end{eqnarray}
where $\langle\cdots\rangle$ denotes a thermal average with respect to $F_{\rm GL}$, and in the case of temperature derivatives, we have omitted some sub-leading terms.  In the case of $h$ derivatives of the free energy, $h$ and $\delta r$ dependences appear in $\langle \varphi \rangle$ but not in their expansion coefficients. In the present case of $s_j$ derivatives, however, the expansion parameters $a^{(i)}_{s_j}$ with $i=1,2,\cdots$ depend on the original model parameters $\vec{s}$ as 
\begin{eqnarray}
  a^{(1)}_{s_j}&=&-\frac{\partial h}{\partial s_j}
  +\frac{\partial u}{\partial s_j}\frac{1}{N_c}
  \int\!d{\bf x}\,\langle (\varphi({\bf x})-\langle\varphi\rangle)^3\rangle
  \label{eq:a1}\\
  a^{(2)}_{s_j}&=&\frac{1}{2}\frac{\partial r}{\partial s_j}
  +3\frac{\partial u}{\partial s_j}\frac{1}{N_c}
  \int\!d{\bf x}\,\langle(\varphi({\bf x})-\langle\varphi\rangle)^2\rangle
  \label{eq:a2}
\end{eqnarray}
They satisfy $\partial a^{(\ell)}_{s_j}/\partial s_{j'}=\partial a^{(\ell)}_{s_j'}/\partial s_j$, which is referred to as $b^{(\ell)}$. In particular, at the Mott transition boundary or a metal-insulator crossover $h=0$, we obtain
\begin{subequations}
  \begin{eqnarray}
    a^{(1)}_{s_j}&=&-\frac{\partial h}{\partial s_j}
    \label{eq:d1_MIT}\\
    b^{(1)}_{s_j,s_{j'}}&=&-\frac{\partial^2 h}{\partial s_j \partial s_{j'}}.
    \label{eq:b1_MIT}
  \end{eqnarray}
\end{subequations}

\subsection{Second derivatives}
\label{app:derivative:second}

Here we give an expression of the second derivative with respect to $s_j$ and $s_{j'}$;
\begin{eqnarray}
  &&-\frac{1}{N_c}\frac{\partial^2 F}{\partial s_j \partial s_{j'}}
  \nonumber\\
  &=&\frac{1}{N_c}\langle(\frac{\partial F_{\rm GL}}{\partial s_j}-\langle\frac{\partial F_{\rm GL}}{\partial s_j}\rangle)(\frac{\partial F_{\rm GL}}{\partial s_{j'}}-\langle\frac{\partial F_{\rm GL}}{\partial s_{j'}}\rangle)+\frac{\partial^2 F_{\rm GL}}{\partial s_j \partial s_{j'}}\rangle
  \nonumber\\
  &\approx&
  \frac{\partial h}{\partial s_j} \frac{\partial h}{\partial s_{j'}}\chi
  -(\frac{\partial h}{\partial s_j}   \frac{\partial r}{\partial s_{j'}}
  + \frac{\partial h}{\partial s_{j'}}\frac{\partial r}{\partial s_j})\chi_3
  +\frac{\partial r}{\partial s_j}\frac{\partial r}{\partial s_{j'}}\chi_4+\cdots
  \nonumber\\
  &&\!\!\!+b^{(1)}\langle\varphi\rangle
  +b^{(2)}\langle\varphi\rangle^2+\cdots,
  \label{eq:chi_Dn}
\end{eqnarray}
where
\begin{equation}
  \chi_3\equiv \frac{1}{N}\frac{\partial^2 F}{\partial h \partial r}
  \label{eq:chi_3}
\end{equation}
vanishes at criticality and
\begin{equation}
  \chi_4\equiv -\frac{1}{N}\frac{\partial^2 F}{\partial r^2}
  \label{eq:chi_4}
\end{equation}
has a critical divergence characterized by
\begin{subequations}
\begin{eqnarray}
  \chi_4|_{h=0,r} &\propto& |r-r_{\rm cr}|^{-\alpha}
  \label{eq:chi_4_alpha}\\
  \chi_4|_{h,r=r_{\rm cr}} &\propto& |h|^{-\alpha/\beta\delta}
  \label{eq:chi_4_gamma}
\end{eqnarray}
\end{subequations}
with the specific heat exponent $\alpha=2-d\nu$ and hence is subdominant for $\alpha<\gamma$.

\section{Derivation of Renormalization Group Equations}
\label{app:RG}

Here, we derive the renormalization group equations given in Eq.~(\ref{eq:RG}). First, we add to the GL free energy Eq.~(\ref{eq:F_GL_interplay}) a third-order coupling term $(g_\varphi/6)\int d{\bf x}\varphi({\bf x})^3$ and define the corresponding dimensionless coupling constant $\tilde{g}_\varphi\equiv\sqrt{I_d}g_\varphi$. Then, the RG equations are obtained up to the one-loop order as follows;
\begin{subequations}
  \begin{eqnarray}
    \frac{d\tilde{g}_\varphi}{d\lambda}
    &=&[-\varepsilon_3
    +9\tilde{u}_{\varphi\varphi}-\tilde{g}_\varphi^2] \tilde{g}_\varphi
    +[\tilde{u}_{\varphi\phi} - m\tilde{g}^2]\tilde{g}
    \label{eq:app:RG:g_varphi}\\
    \frac{d\tilde{g}}{d\lambda}
    &=&[-\varepsilon_3
    +(m+2)\tilde{u}_{\varphi\phi}
    +(m+8)\tilde{u}_{\phi\phi}
    -\tilde{g}^2-\tilde{g}_\varphi\tilde{g}]\tilde{g}
    \nonumber\\
    \label{eq:app:RG:g}\\
    \frac{d\tilde{u}_{\varphi\varphi}}{d\lambda}
    &=&-\varepsilon \tilde{u}_{\varphi\varphi}
    +9\tilde{u}_{\varphi\varphi}^2+m\tilde{u}_{\varphi\phi}^2
    +\frac{1}{2}(\tilde{g}_\varphi^4+m\tilde{g}^4)
    \label{eq:app:RG:u_varphivarphi}\\
    \frac{d\tilde{u}_{\phi\phi}}{d\lambda}
    &=&-\varepsilon\tilde{u}_{\phi\phi}+(m+8)\tilde{u}_{\phi\phi}^2
    +\tilde{u}_{\varphi\phi}^2+\tilde{g}^4
    \label{eq:app:RG:u_phiphi}\\
    \frac{d\tilde{u}_{\varphi\phi}}{d\lambda}
    &=&-\varepsilon \tilde{u}_{\varphi\phi}
    +[3\tilde{u}_{\varphi\varphi}+(m+2)\tilde{u}_{\phi\phi}
      +4\tilde{u}_{\varphi\phi}]\tilde{u}_{\varphi\phi}
    \nonumber\\
    &&{}+\tilde{g}^4+\tilde{g}_\varphi\tilde{g}^3+\tilde{g}_\varphi^2\tilde{g}^2.
    \label{eq:app:RG:u_varphiphi}
  \end{eqnarray}
  \label{eq:app:RG}
\end{subequations}

Next, we impose $\tilde{g}_\varphi=0$ at each value of $\lambda$ by renormalizing $\tilde{\varphi}(x)\equiv\varphi(x)\Lambda^{(d-2)/2}$ as $\tilde{\varphi}(x)\to e^{(d-2)\delta\lambda/2}\tilde{\varphi}(x)-\delta\tilde{g}_\varphi/(6u_{\varphi\varphi})$.
 This introduces an additional term $\tilde{u}_{\varphi\phi}/(6\tilde{u}_{\varphi\varphi})\cdot d\tilde{g}_\varphi/d\lambda$ to the RG Eq.~(\ref{eq:app:RG:g}).
Finally, substituting $\tilde{g}_\varphi=0$ into the RG equations, we obtain Eqs.~(\ref{eq:RG}).

\end{appendix}


\begin{thebibliography}{9}
\bibitem{Mott}
  N.F. Mott, {\it Metal-Insulator Transitions} (Taylor and Francis, London/Philadelphia, 1990).
\bibitem{Slater}
  J.C. Slater, \citePR{82}{1951}{538}.
\bibitem{Hubbard}
  J. Hubbard, \citePRSA{276}{1963}{238}; \citeIBID{281}{1964}{401}.
\bibitem{RMP_Kadanoff}
  For a combined review of theories and experiments on liquid-gas phase transitions,
  see L.P. Kadanoff, {\it et al}., \citeRMP{39}{1967}{395}.
\bibitem{Cyrot72}
  M. Cyrot, \citeJPF{}{33}{1972}{125}.
\bibitem{Castellani79}
  C. Castellani, C. Di Castro, D. Feinberg and J. Ranninger, \citePRL{43}{1979}{1957}.
\bibitem{RMP_ImadaFujimoriTokura}For a review, see
  M. Imada, A. Fujimori and Y. Tokura, \citeRMP{70}{1998}{1039}.
\bibitem{OnodaNagaosa03}
  For a summary of recent experimental findings on the Mott transition in $\kappa$-(BEDT-TTF)$_2$X, see S. Onoda and N. Nagaosa, \citeJPSJ{72}{2003}{2445} and references therein. 
\bibitem{BednortzMuller86}
  J.G. Bednortz and K.A. M\"{u}ller, Z. Phys. B {\bf 64}, 189 (1986).
\bibitem{Anderson73}
  P.W. Anderson, Mater. Res. Bull. {\bf 8}, 153 (1973).
\bibitem{RVB}
  P.W. Anderson, Science {\bf 235}, 1196 (1987).
\bibitem{Kagawa_unpublished}
  F. Kagawa, T. Itou, K. Miyagawa, and K. Kanoda, unpublished. See also F. Kagawa, T. Itou, K. Miyagawa, and K. Kanoda, \citePRB{69}{2004}{064511}.
\bibitem{ShimizuMiyagawaKanodaMaesatoSaito03}
  Y. Shimizu, K. Miyagawa, K. Kanoda, M. Maesato, and G. Saito, unpublished (cond-mat/0307483).
\bibitem{Tokura_MIT_pyrochlore}
  Y. Tokura, private communications; Y. Taguchi, K. Ohgushi, and Y. Tokura, \citePRB{65}{2002}{115102}.
\bibitem{BrinkmanRice70}
  W.F. Brinkman and T.M. Rice, \citePRB{2}{1970}{4302}.
\bibitem{RMP_DMFT}W. Metzner and D. Vollhardt, \citePRL{62}{1989}{324};
  A. Georges {\it et al.}, \citeRMP{68}{1996}{13}.
\bibitem{KotliarRuckenstein86}
  G. Kotliar and A.E. Ruckenstein, \citePRL{57}{1986}{1362}.
\bibitem{Stanley87}
  H.E. Stanley, {\it Introduction to Phase Transitions and Critical Phenomena} 
  (Oxford Univ. Press, New York, 1987).
\bibitem{GL_DMFT}
  G. Kotliar, \citeEPJB{11}{1999}{27};
  G. Kotliar, E. Lange, and M.J. Rozenberg, \citePRL{84}{2000}{5180}.
\bibitem{DMFA}
  Th. Pruschke {\it et al}., Adv. in Phys. {\bf 42}, 187 (1995).
\bibitem{Goldenfeld}
  N. Goldenfeld, {\it Lectures on Phase Transitions and the Renormalization Group} (Addison-wesley, Boston, 1992).
\bibitem{Science_Limelette03}
  P. Limelette {\it et al.}, Sicence {\bf 302}, 89 (2003).
\bibitem{KotliarMurthyRozenberg02}
  G. Kotliar, S. Murthy, and M.J. Rozenberg, \citePRL{89}{2002}{046401}.
\bibitem{CPM}
  S. Onoda and M. Imada, \citePRB{67}{2003}{161102}.
\bibitem{RozenbergChitraKotliar99}
  M.J. Rozenberg {\it et al.}, \citePRL{83}{1999}{3498}.
\bibitem{DCA}
  Th. Maier {\it et al.}, \citeEPJB{13}{2000}{613};
  M. Jarrell {\it et al.}, \citePRB{64}{2001}{195130}.
\bibitem{CDMFT}
  G. Kotliar {\it et al.}, \citePRL{87}{2001}{186401}.
\bibitem{OPM}S. Onoda and M. Imada, \citeJPSJ{70}{2001}{632};
  \citeJPSJ{70}{2001}{3398};
  \citeJPCS{63}{2002}{2225}. 
\bibitem{MatsumotoMancini97}H. Matsumoto and F. Mancini, \citePRB{55}{1997}{2095}.
\bibitem{AvellaManciniMunzner01}
  A. Avella, F. Mancini, and R. M\"{u}nzner, \citePRB{63}{2001}{245117}.
\bibitem{Roth69}
  L.M. Roth, \citePR{184}{1969}{451}.
\bibitem{OP}S. Nakajima, \citePTP{20}{1958}{948};
  R. Zwanzig, {\it Lectures in Theoretical Physics}, Vol. 3, (Interscience, New York, 1961);
  H. Mori, \citePTP{33}{1965}{423}; \citePTP{34}{1965}{399}.
\bibitem{note_superconductivity}
Actually, if superconductivity occurs, namely, the BCS self-energy parts remain finite, then we must extend the above representation of the Dyson equation into that containing charge off-diagonal elements, like the Nambu representation. This is straightforwardly done by taking a $2M \times 2M$ matrix representation, for instance, which is spanned by $(c_{\mibs{k},\sigma},c^\dagger_{\mibs{k},\sigma})$ with $\sigma=1,\cdots,M$ for each momentum ${\bf k}$. These additional elements can be absorbed into the component degrees of freedom. Therefore, one can implicitly take this enlarged representation in discussing a superconducting state. 
\bibitem{noteDCA}
  In the equation of the DCA~\cite{DCA}, the ${\bf k}$ off-diagonal components in Eq.~(\ref{eq:Sigmac2Sigma}) are assumed to be absent due to a translational symmetry.
\bibitem{Moriya}
  T. Moriya, {\it Spin Fluctuations in Itinerant Electron Magnetism} (Springer-Verlag, Berlin, 1985).
\bibitem{Hertz76}
  J.A. Hertz, \citePRB{14}{1976}{1165}.
\bibitem{Millis93}
  A.J. Millis, \citePRB{48}{1993}{7183}.
\bibitem{Ma}
  S.K. Ma, {\it Modern Theory of Critical Phenomena} (Addison-Wesley, Redwood City, 1976).
\bibitem{ChakravartyHalperinNelson89}
  S. Chakravarty, B.I. Halperin, and D.R. Nelson, \citePRB{39}{1989}{2344}.
\bibitem{note_phasediagram}
  In the schematic phase diagram proposed by Kotliar {\it et al}~\cite{KotliarMurthyRozenberg02}, they did not clarify that in the $(U,\mu,T)$ space, the shape of the Mott phase boundary is free from singularity even at the curve where a bandwidth-control Mott transition but not a filling-control occurs.
\bibitem{OnodaImada_jmmm}
  S. Onoda and M. Imada, J. Mag. Mag. Mat. {\bf 272-276}, Suppl. 1, E275 (2004).
\bibitem{Anzai86}
  S. Anzai, M. Matoba, M. Hattori, and H. Sakamoto, \citeJPSJ{55}{1986}{2531}.
\bibitem{MenthRemeika70}
  Early measurements of the uniform spin susceptibility in V$_2$O$_3$ [A. Menth and J.P. Remeika, \citePRB{2}{1970}{3756}] revealed that a metal-insulator crossover was accompanied by that of the Curie constant, which increased from a metallic side to an insulating.
\bibitem{GossardMcWhanRemeika70}
  A.C. Gossard, D.B. McWhan, and J.P. Remeika, \citePRB{2}{1970}{3762}.
\bibitem{Slichter}
  C.P. Slichter, {\it Principles of Magnetic Resonance}, 3rd ed. (Springer-Verlag, Berlin, 1989). 
\bibitem{Fournier}
  D. Fournier, M. Poirier, M. Castonguay, and K.D. Truong, \citePRL{90}{2003}{127002}.
\bibitem{Amit}
  For instance, see D.J. Amit, {\it Field Theory, the Renormalization Group, and Critical Phenomena}, 2nd ed. (World Scientific, Singapore, 1984).
\end{thebibliography}
\end{document}